\documentstyle[12pt]{article}
\begin{document}

\begin{titlepage}

\vspace*{-2cm}

\hspace*{\fill} CGPG-96/1-7\\
\vspace{.5cm}

\begin{centering}

{\huge Isomorphisms between the Batalin-Vilkovisky antibracket and
the Poisson bracket} 

\vspace{.5cm}
{\large Glenn Barnich$^{1,*}$ and  
Marc Henneaux$^{2,**}$}\\
\vspace{1cm}
$^1$Center for Gravitational Physics and Geometry, 
The Pennsylvania State University,
104 Davey Laboratory, University Park, PA 16802 \\
\vspace{.5cm}
$^2$Facult\'e des Sciences, Universit\'e Libre de Bruxelles,
Boulevard du Triomphe, Campus Plaine C.P. 231, 
B-1050 Bruxelles, Belgium

\vspace{.5cm}

\begin{abstract}
One may introduce at least three different Lie algebras in any
Lagrangian field
theory : (i) the Lie algebra of local BRST cohomology classes
equipped with the odd Batalin-Vilkovisky antibracket, which has attracted
considerable interest recently~; (ii) the Lie algebra of local
conserved currents equipped with the Dickey bracket~; and (iii) the
Lie algebra of conserved, integrated charges equipped with the 
Poisson bracket. We
show in this paper that the subalgebra  of (i) in ghost number $-1$
and the other two algebras are isomorphic for a field
theory without gauge invariance. We also prove that, in the presence
of a gauge freedom, (ii) is still isomorphic to the subalgebra of (i) in 
ghost
number $-1$, while (iii) is isomorphic to the quotient of (ii) by the
ideal of currents without charge. In ghost number different from $-1$,
a more detailed analysis of the
local BRST cohomology classes in the Hamiltonian formalism allows 
one to
prove an isomorphism theorem between the antibracket and the extended 
Poisson bracket of Batalin, Fradkin and Vilkovisky. 

\end{abstract}

\end{centering}

\vspace{1.5cm}

{\footnotesize \hspace{-0.6cm}($^*$)Aspirant au Fonds National de la
Recherche Scientifique (Belgium).\\
($^{**}$)Also at Centro de Estudios
Cient\'\i ficos de Santiago, Chile.}
\end{titlepage}

\pagebreak
\def\qed{\hbox{${\vcenter{\vbox{                         
   \hrule height 0.4pt\hbox{\vrule width 0.4pt height 6pt
   \kern5pt\vrule width 0.4pt}\hrule height 0.4pt}}}$}}
\newtheorem{theorem}{Theorem}[sectionc]
\newtheorem{lemma}{Lemma}
\newtheorem{definition}{Definition}
\newtheorem{corollary}{Corollary}
\newcommand{\proof}[1]{{\bf Proof.} #1~$\qed$.}
\renewcommand{\theequation}{\thesection.\arabic{equation}}
\renewcommand{\thetheorem}{\thesection.\arabic{theorem}}
\renewcommand{\thelemma}{\thesection.\arabic{lemma}}
\renewcommand{\thecorollary}{\thesection.\arabic{corollary}}

\section{Introduction}
\setcounter{equation}{0}
The first appearance of an antibracket in the context of Lagrangian
field theories can be traced back to the study of the renormalization of
Yang-Mills theories when the Ward identities are expressed in terms 
of the generating functional for one particle irreducible
proper vertices \cite{Zinn-Justin}. This antibracket
has been developped and generalized in the work of Batalin and Vilkovisky
\cite{Batalin} on Lagrangian quantization methods for generic gauge
theories. The Batalin-Vilkovisky formalism and the antibracket play
for instance a
fundamental role in the covariant formulation of string field theory
\cite{Bochicchio}. 
It is therefore of interest to gain a better understanding of
the physical significance of this antibracket.

We relate in this paper the Batalin-Vilkovisky antibracket 
at ghost number minus one both 
to the
bracket introduced by Dickey \cite{Dickey} in the space of local
currents, and to the Poisson bracket of conserved charges. More
generally, we relate the Batalin-Vilkovisky antibracket for arbitrary
values of the ghost number to the extended Poisson
bracket appearing in the Hamiltonian formulation of the BRST theory 
\cite{BFV,Henneaux}.   

The paper is organized as follows. In the next section, we review the
Batalin-Vilkovisky construction and show that the Batalin-Vilkovisky
antibracket naturally 
induces a well defined odd Lie bracket $\{\cdot,\cdot\}$ in 
the cohomology classes $H^{*,n}(s|d)$ of
the BRST differential $s$ modulo the exterior spacetime differential
$d$ in form degree $n$. 
The algebra $(H^{*,n}(s|d),\{\cdot,\cdot\})$ possesses a subalgebra
${\cal S}$,
namely $(H^{-1,n}(s|d),\{\cdot,\cdot\})$. 

We then define the Dickey
algebra of conserved currents $j^\mu$ (section 3) 
and show that it possesses an ideal, 
namely the ideal $I$ of non trivial 
conserved currents for which the charge
$Q=\int d^{n-1}x j^0$ is zero on-shell. Such currents are trivial
(i.e., on-shell equal to identically conserved currents) when there is
no gauge freedom, so that $I$ is effectively zero in that case. They
may however be non trivial otherwise. We introduce furthermore 
the Lie algebra of integrated conserved charges equipped with the 
covariant Poisson bracket induced by the Dickey bracket. 

Isomorphism theorems between ${\cal S}$ and the other two Lie algebras
in the
case of non degenerate field theories are proved in section 4. The
modification of these theorems for gauge theories are 
discussed in section 5. More precisely, we show that ${\cal S}$ is
still isomorphic to the Dickey algebra, but this algebra itself is now 
isomorphic to the Lie algebra of conserved charges only after taking
the quotient by the ideal $I$. 

In section 6, we investigate the antibracket map for arbitrary ghost
number. In order to do so, we go to the extended Hamiltonian
formalism and use the fact that the local
BRST cohomology group and the associated antibracket map are invariant
under this change of description of the theory. The advantage of the
Hamiltonian formulation is that the equations of motion are in normal
form, which allows one to control the antifield dependence of the local
BRST cohomology classes. We show that it is always
possible to 
choose representatives which are at most linear in the antifields of
the Hamiltonian description. This allows one to get the general
relationship between the antibracket map and the extended Poisson
bracket map of the Hamiltonian BRST formalism. 

By applying these
results to the case of ghost number $-1$, we find in particular that
$I$ is an abelian subalgebra and corresponds to a subspace of 
the characteristic
cohomology associated with the Hamiltonian constraint surface.

\section{The antibracket map induced in local BRST cohomology}
\setcounter{equation}{0}
In the Batalin-Vilkovisky formalism for gauge theories, which we
consider for notational simplicity to be irreducible, one introduces,
besides the original fields $\phi^i$ of ghost number $0$
and the ghosts $C^\alpha$ of ghost number $1$ related
to the gauge invariance, the corresponding antifields $\phi^*_i$ and 
$C^*_\alpha$ of opposite Grassmann parity and ghost number $-1$ and
$-2$ respectively \cite{Batalin,Henneaux}. 
It is natural to define an antibracket by
declaring that the fields $\phi^A\equiv(\phi^i,C^\alpha)$ and
antifields $\phi^*_A$ are conjugate~:
\begin{eqnarray}
(\phi^A(x),\phi^*_B(y))=\delta^A_B \delta^n(x-y)
\end{eqnarray}
The antibracket is then given for arbitrary functionals 
$A_1$ and $A_2$ by
\begin{eqnarray} 
(A_1,A_2)=\int d^nx({\delta^RA_1\over\delta\phi^A(x)}
{\delta^LA_2\over\delta\phi^*_A(x)}
-{\delta^RA_1\over\delta\phi^*_A(x)}
{\delta^LA_2\over\delta\phi^A(x)}).
\end{eqnarray}
The central goal of the formalism is the construction of a proper
solution to the master equation 
\begin{eqnarray}
(S,S)=0. 
\end{eqnarray}
The functional $S$ is required to start 
like the classical action $S_0$, to which one couples
through the antifields the gauge transformations with the gauge parameters
replaced by the ghosts : 
\begin{eqnarray}
S=\int d^nx \hat {\cal L}=\int d^n x \hat {\cal L}_0 +\phi^*_i
R^i_\alpha C^\alpha +\dots .
\end{eqnarray}
The BRST symmetry is canonically generated in the antibracket 
through the equation~:
\begin{eqnarray}
s=(S,\cdot).
\end{eqnarray}

In order to analyze the properties of the 
antibracket, it is necessary to have a more
precise definition of the functionals to which it applies. We will
consider in the following only local functionals. A local functional
\begin{eqnarray}
A[z^a(x)]=\int_{X} d^nx\ \hat a[z^a],\  z^a(x)\rightarrow0\ for\
x\rightarrow \partial X
\end{eqnarray}
is defined as the integral over an orientable domain $X$ of
spacetime $M^n$ 
of a local function $\hat a[z^a]$, 
i.e., a function\footnote{We will not be too precise about 
the nature of the field dependence of the local functions  
(polynomiality or smooth dependence). Similarily, we will not specify
whether one should consider polynomials or infinite formal 
series in the antifields and 
their derivatives \cite{BBH}, since most aspects we will consider are 
really independent of these considerations. For simplicity, we 
will assume however that all the fields live on a star-shaped space.}
of $x^\mu$, the fields and
antifields $z^a\equiv (\phi^A,\phi^*_A)$ and their
derivatives up to some finite order, evaluated for field and antifield
histories $z^a(x)$ which appropriately 
vanish at the boundary $\partial X$. Note
that $X$ can
be all of Minkowski space $M^n$, and that a local function corresponds to
a function on the finite dimensional ``jet-space''
$M^n\times V^k$ with coordinates 
$x^\mu,\partial_{(\nu)}z^a, |\nu|\leq k$ (see appendix A for more details). 
The space of local functionals so defined can be proved
(see for instance \cite{Olver,Henneaux}) to be isomorphic
to the space of equivalence classes of local functions $\hat a$ modulo total
divergences $\partial_\mu j^\mu$, for some 
arbitrary local current $j^\mu$. The
total derivative $\partial_\mu$ is defined in multiindex notation by  
\begin{eqnarray}
\partial_\mu={\partial^L\over\partial x^\mu}+\partial_{\mu(\nu)}
z^a{\partial^L\over\partial(\partial_{(\nu)}z^a) }.
\end{eqnarray}
One can furthermore prove that a local
function is a total divergence if and only if its Euler-Lagrange
derivatives vanish (see e.g. \cite{Olver}).

Turning to form notations, $\hat a\rightarrow a=d^nx \hat a$
and introducing the spacetime exterior
derivative $d=dx^\mu\partial_\mu$, the space of local functions can be
identified with the cohomology group $H^n(d)$ of the differential $d$
in form degree $n$ in the space of local, form valued functions.

It is easy to verify that the antibracket of two local functionals is
also a local functional. Thus the antibracket induces a well defined
map in the cohomology group $H^n(d)$,
\begin{eqnarray}
\{\cdot,\cdot\} : H^n(d)\times H^{n}(d)\longrightarrow H^{n}(d)
\label{2.8}
\end{eqnarray}
This bilinear map enherits from the antibracket the property of being
a true, odd, Lie bracket. If we denote by $[a]$ the cohomological
class of the $n$-form $a$ in $H^n(d)$, one may view the antibracket in
$H^n(d)$ as arising from a local antibracket in the space of local
functions defined as follows,
\begin{eqnarray}
\{\hat a_1,\hat a_2\}={\delta^R\hat a_1\over\delta\phi^A}
{\delta^L\hat a_2\over\delta\phi^*_A}
-{\delta^R\hat a_1\over\delta\phi^*_A}
{\delta^L\hat a_2\over\delta\phi^A}\label{la}\\
\{[a_1],[a_2]\}=[d^nx \{\hat a_1, \hat a_2\}].
\end{eqnarray}
In (\ref{la}), $\delta/\delta\phi^A$ is the Euler-Lagrange derivative
defined by 
\begin{eqnarray}
{\delta\over\delta\phi^A}=
(-\partial)_{(\nu}){\partial\over\partial(\partial_{(\nu)}\phi^A)}, 
\end{eqnarray}
with $(-\partial)_{(\nu)}=(-)^{|\nu|}\partial_{(\nu)}$. 
While the bracket (\ref{2.8}) in $H^n(d)$ is a true bracket, the local
antibracket (\ref{la}) in the space of local functions
is graded symmetric, but satisfies the graded
Leibnitz rule and Jacobi identity only up to total divergences (see
Appendix B).

It is clear that the antibracket for the integrands that gives rise
to the antibracket in $H^n(d)$ is not unique, but
expressions differing from the one in (\ref{la}) by a total
divergence are also
admissible. This is the case for instance for the following expression 
(see appendix B),
\begin{eqnarray}
\{\hat a_1,\hat a_2\}_{alt}=\partial_{(\nu)}({\delta^R\hat
a_1\over\delta\phi^A})
{\partial^L\hat a_2\over\partial(\partial_{(\nu)}\phi^*_A)}
-\partial_{(\nu)}({\delta^R\hat
a_1\over\delta\phi^*_A}){\partial^L\hat a_2
\over\partial(\partial_{(\nu)}\phi^A)},\label{alt}
\end{eqnarray}
which satifies a graded Leibnitz rule in the second argument, but is
only graded symmetric up to a total divergence. [There is no
expression for the local antibracket in the space of local functions
that satisfies strictly all the
properties of an ordinary odd Lie bracket, without extra divergences.] 

In the Batalin-Vilkovisky formalism, one introduces additional
fields, the ghosts and antifields. Quantities of direct physical
interest are recovered by considering 
the cohomology classes of the BRST differential
$s$. The identification of local functionals with the cohomology group
$H^n(d)$ implies that the BRST cohomology for local functionals is
given by $H^*(s,H^n(d))$. This last group is 
isomorphic to the relative cohomology group $H^{*,n}(s|d)$ of $s$
modulo $d$ in form degree $n$ evaluated in the space of form valued
local functions. Due to the fact that the BRST symmetry acting on a 
local function is canonically generated through the formula
\begin{eqnarray}
s \hat a=\{\hat {\cal L}, \hat a\}_{alt},
\end{eqnarray} 
it is straightforward\footnote{One uses the facts that
(i) $\{\cdot,\cdot\}_{alt}$ differs from $\{\cdot,\cdot\}$ by a total
divergence, (ii) that $\{\cdot,\cdot\}$
satisfies the graded Jacobi up to a total divergence, and (iii) that
Euler-Lagrange derivatives annihilate total divergences.} 
to verify that the local antibracket induces a
well defined odd Lie bracket in the relative cohomology group of $s$
modulo $d$ :
\begin{eqnarray}
\{\cdot,\cdot\} : H^{n,g_1}(s|d)\times H^{n,g_2}(s|d)\longrightarrow 
H^{n,g_1+g_2+1}(s|d)\nonumber\\
\{[a_1],[a_2]\}=[d^nx \{\hat a_1, \hat a_2\}]\label{abm}.
\end{eqnarray}
An inspection of the
various possible cases shows that it is only for ghost number $-1$
that this map associates to two cohomology classes a
cohomology class of the same type, i.e., of same ghost number. The
subspace $H^{-1,n}(s|d)$ equipped with the antibracket defines a
subalgebra of $H^{n,*}(s|d)$ which we denote by ${\cal S}$,
\begin{eqnarray}
{\cal S}=(H^{-1,n}(s|d),\{\cdot,\cdot\}).
\end{eqnarray}

\section{The Dickey bracket}
\setcounter{equation}{0}
Let $\Sigma_k$ be the stationary surface, i.e., the surface defined by
the equations 
\begin{eqnarray}
\partial_{(\lambda)} (\delta \hat {\cal L}_0/\delta
\phi^i)=0, 
\end{eqnarray}
(with $|\lambda|\leq k-2$ for second order equations) 
in the spaces $M\times F^k$ with
coordinates $x^\mu,\partial_{(\mu)}\phi^i$, $|\mu|\leq k$.

The vector space of (equivalence classes of) inequivalent
Lagrangian conservation laws is defined by 
\begin{eqnarray}
\{j^\mu, \partial_\mu j^\mu \approx 0,{\rm modulo\ the\ identification\ } 
j^\mu\sim|_{\Sigma} j^\mu +\partial_\nu
S^{[\nu\mu]}\}\label{toll},
\end{eqnarray} 
where the $j^\mu$ are local functions. 
In form notations, we
get equivalence classes $[j]$ of $n-1$ forms whose
pull-back to the stationary surface is
$d$-closed,
where two such forms have to be identified if they differ by the
exterior derivative of an $n-2$ form on the stationary surface :
\begin{eqnarray}
[j]\in H^{n-1}(d^*,\Omega(\Sigma)). 
\end{eqnarray}
Inequivalent conserved currents belong, by definition,
to the so called characteristic
cohomology of the stationary surface in form degree $n-1$. 

The standard 
regularity conditions are that locally in the jet-space, 
the equations ${\delta \hat {\cal L}_0/\delta
\phi^i}$ and
their derivatives can be split into two groups, the ``independent
equations'' which can be taken locally as a new coordinate system on the
jet-space 
replacing some of the fields and their derivatives,
and the ``dependent equations'' which hold as a
consequence of the independent ones. One then can prove 
\cite{Olver,Henneaux} that a function
which vanishes on the stationary surface can be written as a linear
combination of the equations defining the surface,
hence 
\begin{eqnarray}
\partial_\mu j^\mu=X^{i(\lambda)}
\partial_{(\lambda)} {\delta \hat {\cal L}_0\over\delta
\phi^i}\label{bol}
\end{eqnarray}
for some local functions $X^{i(\lambda)}$. This equation does not
determine $X^{i(\lambda)}$ completely, one is for instance free
to add functions of the form 
\begin{eqnarray}
Y^{i(\lambda)j(\nu)}\partial_{(\nu)} 
{\delta \hat {\cal L}_0\over\delta\phi^j}
\end{eqnarray}
with $Y$ antisymmetric
under the exchange of the pairs $i(\lambda)$ 
and $j(\nu)$\footnote{This exhausts the arbitrariness of the
functions $X^{i(\lambda)}$ only in the case
where the equations and their derivatives are independent
\cite{Olver,Henneaux}~; in the
general case, one has to take care also of the Noether identities, as
shown below.}.

The characteristic \cite{Olver,Dickey} 
of the equivalence class of conservation laws
described by
$[j]$ is defined by the
equivalence class of local functions of the form  
$X^i=(-\partial)_{(\lambda)}X^{i(\lambda)}$, where two sets $X^i$'s of local
functions   
have to be identified if they differ by a function of the form 
\begin{eqnarray}
(-\partial_{(\lambda)})[Y^{i(\lambda)j(\nu)}\partial_{(\nu)} 
{\delta \hat {\cal L}_0\over\delta\phi^j}].\label{sup}
\end{eqnarray}
It is straightforward to verify that the characteristic does
not depend on the choice of the representative for $j^\mu$.
Let $\delta_X$ be the evolutionary vector field defined by $X^i$~:
\begin{eqnarray}
\delta_X=\partial_{(\lambda)}X^i  
{\partial\over\partial(\partial_{(\lambda)}\phi^i)}.
\end{eqnarray} 
Note that $X^i$ and $\delta_X$ satisfy the equations 
\begin{eqnarray}
X^i{\delta \hat {\cal L}_0\over\delta
\phi^i}=\partial_\mu j^{\prime\mu},\ \delta_X \hat {\cal L}_0=\partial_\mu
j^{\prime\prime\mu},
\end{eqnarray}
with $j^{\prime\mu},j^{\prime\prime\mu}$ in the same equivalence class
as $j$. This means that the characteristics define 
variational symmetries, i.e., symmetries of the
action. In the non degenerate case, one can then prove directly 
that there is a
one to one correspondence between inequivalent symmetries of the
action and inequivalent conservation laws (Noether's theorem) 
\cite{Olver}, 
but we will not do so here because it is also a direct consequence of 
our analysis
in the next section.

The Dickey bracket (\ref{toll}) in the space of inequivalent
conservation laws is defined by \cite{Dickey}
\begin{eqnarray}
\{[j_1],[j_2]\}_D=-[\delta_{X_1} j_2]\label{x}.
\end{eqnarray}
By using properties of the Euler-Lagrange derivatives, one finds
the following equivalent expressions 
(see \cite{Dickey} and Appendix B):
\begin{eqnarray}
\{[j_1],[j_2]\}_D
=[\delta_{X_2}j_1]={1\over2}
[\delta_{X_2}
 j_1-\delta_{X_1}j_2]\nonumber\\
=[-{\tilde{(\nu)}_\mu+1\over|\nu|+1}\partial_{(\nu)}
[\delta_{X_2}({\delta{\cal L}_0\over 
\delta(\partial_{(\nu)\mu}\phi^i)})X^i_{1}-\delta_{X_1}
({\delta{\cal L}_0\over\delta(\partial_{(\nu)\mu}\phi^i)})
X^i_{2}]\nonumber\\
{1\over (n-1)!}\varepsilon_{\mu_1\dots\mu_n}dx^{\mu_2}\dots
dx^{\mu_n}],\label{br1}
\end{eqnarray}
where ${(\nu)}_\mu$ denotes the number of occurences of $\mu$ 
in the multiindex $(\nu)$. This last expression corresponds to the
contraction of the horizontal $(n-1)-$ and vertical $2-$ form 
\begin{eqnarray}
\Omega=
{1\over (n-1)!}\varepsilon_{\mu_1\dots\mu_n}dx^{\mu_2}\dots
dx^{\mu_n}{\tilde\mu+1\over|\nu|+1}\partial_{(\nu)}[d_V({\delta{\cal
L}_0\over\delta(\partial_{(\nu)\mu_1}\phi^i)})d_V\phi^i]
\end{eqnarray}
with the evolutionary vector fields $\delta_{X_1}$ and $\delta_{X_2}$.
This formula involves the vertical derivatives and the higher order
Euler operators defined for instance in \cite{Olver,Dickey} (see also
appendix A and B).

Again, in the non degenerate case, one can  
prove directly that the Dickey
bracket is a well defined Lie bracket in the space of inequivalent
conserved currents (see \cite{Dickey})~; namely, it is unambiguous in
the quotient space, antisymmetric and satisfies the Jacobi identity. 
Alternatively, these properties 
follow from the isomorphism theorem proved in the next section.
 
Among the conserved currents, one may distinguish between those for
which $j^0$ is trivial, i.e., of the form $j^0\approx\partial_m
S^{m0}$. The corresponding Noether charge $Q=\int d^{n-1}x j^0$ is
zero on the stationary surface. These currents form an ideal for the
Dickey bracket since $\delta_X j^0$ is trivial if $j^0$ is trivial. We
call this ideal the ideal of ``conserved currents without charge'' and
we denote it by $I$. 

As we shall show in the next section, the ideal $I$ is trivial in the
absence of gauge symmetry. That is, if a conserved current has a
vanishing Noether charge, then, it is trivial, i.e., on-shell equal to
an identically conserved current. But this may not be so in the
presence of gauge freedom, for which there exist non trivial currents
in $I$. 

The third algebra that we shall introduce is the algebra of conserved,
integrated charges, $Q=\int d^{n-1}x j^0$, $\partial_0j^0=-\partial_k
j^k$ for some spatial current $j^k$,
with the identification of two such charges 
if they agree on the stationary surface. By using the
Hamiltonian formalism, one may equip this algebra
with a well defined even
bracket, namely, the standard Hamiltonian Poisson bracket. We denote
this algebra by ${\cal Q}$. It is clear that ${\cal Q}$ is isomorphic
as a vector space to the quotient of the space of conserved currents
by the ideal $I$. We shall prove furthermore
that the Poisson bracket is just the corresponding induced Dickey
bracket. 

\section{Isomorphisms in the case of non-degenerate Lagrangian field theory}
\setcounter{equation}{0}
In the absence of gauge invariance, the only additional fields in the
Batalin-Vilkovisky construction 
besides the original $\phi^i$, which we assume for simplicity to be bosonic, 
are the
antifields $\phi^*_i$. The original action $S[\phi^i]=\int
d^nx\ \hat {\cal L}_0[\phi^i]$ is by itself a proper solution of the master
equation generating the BRST symmetry 
\begin{eqnarray}
s\phi^*_i={\delta \hat {\cal L}_0\over\delta \phi^i},\ s\phi^i=0,\
s\partial_\mu=\partial_\mu s,
\end{eqnarray} 
which reduces to the so called Koszul-Tate differential $\delta$ 
\cite{Henneaux}. In the non degenerate case, the equations of motion 
and their derivatives can be taken
locally as first coordinates in a new coordinate system replacing some
of the fields and their derivatives. One can prove \cite{Henneaux}
that the BRST cohomology in the spaces $C^\infty({\bf R}^n\times
F^k)\times {\bf R}[\partial_{(\nu)}\phi^*_i]$ (with
$|\nu | \leq k-2 $ for second order equations) 
is given by smooth functions defined on the
stationary surface\footnote{One says that the Koszul-Tate 
differential $\delta$ provides a homological resolution of the functions
defined on the stationary surface 
(see also Appendix A).} : $H^0(\delta)\simeq C^\infty({\bf
R}^n\times \Sigma^k)$ and $H^g(\delta)=0, g\neq 0$.

In the new coordinate system, where the equations and their
derivatives are taken as new coordinates, we denote by
$I_0=\{x_a\}$ the set of 
fields and their derivatives needed to complete the coordinate
system. Let us assume that the non degenerate theory is of Cauchy
order $1$, meaning
that $\partial_k x_a\in I_0$ for $k\geq 1$. One can then prove
\cite{BBH} that, apart from $H^{0,n}(s|d)$, which
corresponds to local functionals defined on the stationary surface, the only
non trivial local BRST cohomology classes are in ghost number $-1$ and
form degree $n$. 

By integrations by parts, the representatives of
$H^{-1,n}(s|d)$ can be assumed to be of the characteristic form
\begin{eqnarray}
a=d^nx\ \phi^*_iX^i[\phi^i]
\end{eqnarray} 
for local functions $X^i$. The cocycle condition
reads 
\begin{eqnarray}
{{\delta \hat {\cal L}_0\over\delta \phi^i}}X^i=\partial_\mu j^\mu,
\end{eqnarray}
and implies that the field variation
$\delta\phi^i=X^i$ defines a variational symmetry.
Furthermore, to a trivial representative of
$H^{-1,n}(s|d)$ corresponds a variational symmetry which is given by
an ``antisymmetric'' combination of the equations of motions as in
(\ref{sup})\footnote{A trivial variational symmetry vanishes on the
stationary surface. Under certain assumptions
\cite{BBH}, one can prove that vice-versa every
variational symmetry which vanishes on the stationary surface
corresponds to a trivial representative of $H^{-1,n}(s|d)$, i.e., an
``antisymmetric'' combination of the equations of motion.}. The space
$H^{-1,n}(s|d)$ is accordingly given by inequivalent variational
symmetries or characteristics of inequivalent 
conservation laws\footnote{This is the formalisation in the 
appropriate jet space
of the idea that functions linear in the antifields 
define tangent vectors \cite{Witten},
the physically relevant ones here being those that are ``tangent'' 
to the
stationary surface.}.  

The local antibracket map for such representatives of 
$H^{-1,n}(s|d)$ is given by : 
\begin{eqnarray}
\{[d^nx\ \phi^*_iX^i_1],[d^nx\ \phi^*_i
X^i_2]\}_=[d^nx\ \phi^*_i [X_1,X_2]_L^i]\cr
[X_1,X_2]_L^i={\partial X_1^i\over\partial \phi^j_{(\mu)}}
\partial_{(\mu)}(X^j_2)
-{\partial X_2^i\over\partial \phi^j_{(\mu)}}\partial_{(\mu)}(X^j_1)
=\delta_{X_2} X_1^i-\delta_{X_1} X_2^i.\label{br}
\end{eqnarray}
Hence we find that, in ghost number $-1$, 
the local antibracket map corresponds to
the traditional, even Lie bracket for inequivalent 
variational symmetries under
characteristic form given in \cite{Olver}. Since the Lie bracket for 
evolutionary vector fields is induced by the commutator for vector fields, 
we get~:
\begin{theorem}\label{t1}
The odd Lie algebra ${\cal S}=(H^{-1,n}(s|d),\{\cdot,\cdot\})$ 
is isomorphic to
the algebra of inequivalent variational symmetries equipped with the
bracket induced by the commutator for vector fields.
\end{theorem}

Using the acyclicity of $s=\delta$ \cite{Henneaux} 
at negative ghost numbers and the
triviality of the cohomology of $d$ in form degree $p<n$
($H^p(d)=\delta^p_0{\bf R}$, see e.g. \cite{Olver}), we can easily
prove the isomorphism 
\begin{eqnarray}
H^{-1,n}(\delta|d)\simeq H^{n-1,0}(d|\delta)/\delta^n_1{\bf R}
\label{iso1}. 
\end{eqnarray}
This follows
from a general relationship for relative cohomology groups proved in
\cite{DuboisViolette}. The last space corresponds to the space of 
inequivalent conserved currents. Indeed, the cocycle condition implies
that representatives must be $n-1$-forms which restrict to closed 
forms on the
stationary surface, while the coboundary condition requires two such
currents to be considered as equivalent if they differ on this surface
by the exterior derivative of a $n-2$ form, i.e., the divergence of 
a ``superpotential'' in dual notation, or by a constant in $1$ 
dimension.

The above isomorphism is
explicitly given by associating to a representative $a$ of the first
space the representative $j$ of the second space in the equation
$sa+dj=0$. Furthermore, the antibracket map induces through this
isomorphism a well defined Lie bracket in the space of inequivalent
conserved currents. An explicit calculation (Appendix B) shows 
that the corresponding
bracket is just given by the Dickey bracket. Hence,
\begin{theorem}\label{t2}
The odd Lie algebra ${\cal S}$ is isomorphic to the space of
inequivalent conservation laws equipped with the Dickey bracket.
\end{theorem}
There is no contradiction in the fact that the isomorphism relates an
odd bracket to an even bracket, because 
there is at the same time a shift in
the degree (from odd (-1) to even (0)).

Combining theorems \ref{t1} and \ref{t2}, 
we get the full Noether theorem~:
\begin{corollary}\label{c1}
There is a Lie algebra isomorphism
between inequivalent conservation laws and inequivalent variational
symmetries. 
\end{corollary}

The ideal $I$ of currents of the second set is trivial. Indeed, the
coboundary condition allows us to take all the $j^k$ to depend on the
$x_a$ alone. Because $\partial_k x_a$ depends also on $x_a$ and not on
equations of
motion, one must have $\partial_k j^k=0$ identically, 
which implies that
$j^k=\delta^{n}_2{\bf R}+\partial_m S^{[mk]}$. Hence, the Dickey
algebra and the space of inequivalent, integrated conserved charges
are isomorphic as vector spaces. That the induced Dickey bracket in
the space ${\cal Q}$ corresponds to the Poisson bracket in ${\cal Q}$
in the Hamiltonian formalism is a consequence of
the analysis in section 6. Alternatively, it could be proved directly
along the lines of \cite{Schomblond}, by taking furthermore locality
into account. Hence,
\begin{theorem}\label{t3}
In dimensions
different from $2$, if the theory is of Cauchy order $1$,
the Dickey algebra of inequivalent conserved
currents is isomorphic to the algebra of 
inequivalent conserved charges equipped
with the Poisson bracket.
\end{theorem}

\section{Gauge theories. Ghost number $-1$}
\setcounter{theorem}{0}
\setcounter{equation}{0}
The advantage of the cohomological
reformulation of Noether's theorem in equation (\ref{iso1})
is that one can extend this theorem in a straightforward
way to gauge theories, which are not covered by
the analysis in \cite{Olver,Dickey}. 
One can prove that the subalgebra ${\cal S}$ is
isomorphic to the algebra ${\cal R}=(H_1^n(\delta|d),\{\cdot,\cdot\}_R)$,
where
the cohomology group $H_1^n(\delta|d)$ involves only the original
fields and the antifields,
but no ghosts, $\delta$ being the Koszul-Tate part of $s$ and the
degree of $\delta$ the antighost number, which is minus the ghost
number (for a function that does not involve the ghosts). 
The restricted antibracket map $\{\cdot,\cdot\}_R$ is the
antibracket map restricted to the original fields $\phi^i$ and the
antifields $\phi^*_i$. 

The differential $\delta$ acts non trivially on the antifields of
higher order. In the case of irreducible gauge theories, its action on 
$C^*_a$ is given by 
\begin{eqnarray}
\delta \partial_{(\lambda)}C^*_a
=\partial_{(\lambda)}[R^{+i(\nu)}_a\partial_{(\nu)}\phi^*_i],
\end{eqnarray}
where the operators $R^{+i(\nu)}_a\partial_{(\nu)}$ define the Noether
identities of the theory, i.e., 
\begin{eqnarray}
R^{+i(\nu)}_a\partial_{(\nu)}{\delta \hat {\cal
L}_0\over\delta\phi^i}=0.
\end{eqnarray}
This additional piece maintains $\delta^2=0$ and guarantees that
$\delta$ still defines 
a homological resolution of
the functions defined on the constraint surface, implying for
instance that equation (\ref{iso1}) still holds.

If we still want theorem \ref{t1} to hold, the definition of $\delta$
requires that we change the notion of a trivial variational
symmetry~; they have to correspond to $X^i$'s which are ``antisymmetric''
combinations of the equations of motion up to a gauge transformation
where the gauge parameters are replaced by arbitrary local functions~: 
\begin{eqnarray}
X^i=(-\partial_{(\lambda)})[Y^{i(\lambda)j(\nu)}\partial_{(\nu)} 
{\delta \hat {\cal
L}_0\over\delta\phi^j}]+R^{i(\nu)}_a\partial_{(\nu)}f^a.
\end{eqnarray}
The operators $R^{i(\nu)}_a\partial_{(\nu)}$ are the adjoints of
the operators defining the Noether identities and define the gauge
transformations.

With this modification of the space of inequivalent variational
symmetries, theorems \ref{t1}, \ref{t2} and corollary \ref{c1} hold as
in the case with no gauge invariance.

The ideal $I$ however is not trivial in the case of gauge theories,
because the theory is no longer of Cauchy order $1$. For
instance, the current $j^\mu=F^{0\mu}=(0,F^{0k})$ 
in free Maxwell's theory belongs to $I$ since $\int j^0d^{n-1}x=0$ but
$F^{0k}\neq \partial_m S^{[km]}$ (even weakly). 
Theorem \ref{t3} becomes~:
\begin{theorem}
The Dickey algebra of conserved currents modulo the ideal $I$ is
isomorphic to the algebra of inequivalent conserved charges equipped
with the Poisson bracket.
\end{theorem}
The proof that the induced Dickey bracket is in fact the ordinary
Poisson bracket in the Hamiltonian formalism again follows from the
reasoning given in the next section.
\section{Gauge theories. General analysis}
\setcounter{theorem}{0}
\setcounter{equation}{0}

The previous theorems relate the antibracket and the Poisson bracket
at particular values of the ghost number. In order to fully prove
them, we shall first put them in a more general setting. Indeed, these
theorems can be extended to
arbitrary values of the ghost number. 

To relate the antibracket and the Poisson bracket for all values of
the ghost number,
one first uses the invariance of the 
local BRST cohomology group with 
respect to the introduction of
auxiliary fields and generalized auxiliary fields as shown in
\cite{BBH}. One proves by an analoguous reasoning 
that the same is true for the antibracket map induced in cohomology.
This implies that one can go to the total Hamiltonian
formalism and then to the extended Hamiltonian formalism, which we will 
assume to be local \cite{Henneaux}, and describe
the solution of the master equation in terms of the 
Batalin-Fradkin-Vilkovisky framework. 

Let us recall that in this
framework, a central object is the extended Poisson bracket
$[\cdot,\cdot]_P$ for which
the ghosts $C^a$ and the ghost momenta ${\cal P}_b$ 
are considered as conjugate dynamical variables in
addition to the usual fields and their momenta. One then constructs
out of the constraints, which we assume for simplicity to be
irreducible and first
class, the BRST charge $\Omega=\int d^{n-1}x\ \omega$ which is 
a local functional in space
verifying $[\Omega,\Omega]_P=0$. The Hamiltonian $H=\int d^{n-1}x\ h$ 
verifying 
$[\Omega,H]_P=0$ is also a local functional in space and these two
functionals depend only on the fields 
$\tilde\phi^A\equiv\phi^i,\pi_j,C^a,{\cal
P}_b$ and their spatial derivatives.

The functionals in space are replaced by spatial
functions in the same way as in the spacetime case, 
which leads to a local extended
Poisson bracket $\{\cdot,\cdot\}_P$ defined through
spatial Euler Lagrange derivatives. The BRST charge $\Omega$ generates
the symmetry $s_\omega=\{\omega,\cdot\}_{P,alt}$ where
$\{\cdot,\cdot\}_{P,alt}$ is defined in a way analoguous to
$\{\cdot,\cdot\}_{alt}$ in (\ref{alt}). The local extended Poisson
bracket induces a well defined even Lie bracket, the Poisson bracket
map, in the cohomology group of
$s_\omega$ modulo the spatial exterior derivative $\tilde d$. 

The symmetry 
$s_\omega$ is only a part of
the BRST symmetry which is isomorphic to the BRST symmetry of the
initial Lagrangian system through the elimination of (generalized)
auxiliary fields. The complete BRST symmetry is generated through the 
solution of the master equation in the extended Hamiltonian formalism 
given by 
\cite{Fisch,Henneaux}
\begin{eqnarray}
S_H[\tilde\phi^A,\tilde\phi_A^*]=\int dt d^{n-1}x (-
{1\over 2}\dot{\tilde\phi^A} (\sigma^{-1})_{AB}\tilde\phi^B - h
-\{\tilde\phi^*_A\tilde\phi^A,\omega\}_{P,alt}),\label{meh}
\end{eqnarray} 
where we have introduced the notation 
\begin{eqnarray}
\sigma^{AB}= \left( \begin{array}{cccc}0 & 0
&{\delta_j}^i& 0 \\ 0& 0& 0& -{\delta_{b}}^{a}\\
-{\delta_j}^i&0&0&0\\
0&-{\delta_{b}}^{a}&0&0\end{array}\right).
\end{eqnarray} 
Explicitly, the BRST symmetry $s_H=\{S_H,\cdot\}_{alt}$ reads  
\begin{eqnarray}
s_H=\partial_{(\mu)}({\tilde\delta^R\omega\over\delta\tilde\phi^A})
\sigma^{AB} {\partial^L\over\partial(\partial_{(\mu)}\tilde\phi^B)}
+\partial_{(\mu)}{\cal L}_A
{\partial^L\over\partial(\partial_{(\mu)}\tilde\phi^*_A)},\label{sh}
\end{eqnarray}
where the tilded Euler-Lagrange derivatives are restricted to 
spatial derivatives only and 
\begin{eqnarray}
{\cal L}_A\equiv-\dot{\tilde\phi^B}(\sigma^{-1})_{BA}-
{\tilde\delta^R h\over\delta\tilde\phi^A}-
{\tilde\delta^R\over\delta\tilde\phi^A}
(\{\tilde\phi^*_B\tilde\phi^B,\omega\}_{P,alt})).
\end{eqnarray}
Note that in the proper solution $S_H$ to the master equation in the
extended Hamiltonian formalism, we have made the identification of
minus the antifield $-\lambda^*_a$ of the
Lagrange multiplier for the first class constraints with the  
ghost momenta ${\cal P}_{a}$. This implies that in terms of the new
antifields, the Koszul-Tate part is now associated to the surface
${\cal L}_A(\tilde\phi^*=0)=0$ and not with the gauge invariant,
original, Hamiltonian equations
of motion. The part in resolution degree\footnote{The
resolution degree is the degree associated to the Koszul-Tate
differential \cite{Henneaux}.} $0$ with
respect to the new antifields is given by 
\begin{eqnarray}
\gamma=s^0_\omega-\partial_{(\mu)}{\tilde\delta^R\over\delta\tilde\phi^A}
(\{\tilde\phi^*_B\tilde\phi^B,\omega\}_{P,alt}))\label{so}
\end{eqnarray}
and the BRST differential has no contributions in higher resolution
degree, contrary to what may happen in the old resolution
degree. Here, $s^0_\omega$ is defined by the first term on the right
hand side of equation (\ref{sh}) and coincides with $s_\omega$ when 
acting on a function involving no time derivatives of the fields.
Evaluating the
action of $s^{(0)}_\omega$ on $\phi^i,\pi_j$ and putting to zero 
the ghost momenta ${\cal P}_a$ reproduces the gauge transformations of
these fields with gauge parameters replaced by the ghosts $C^a$.

One then investigates the local BRST cohomology groups $H(s_H|d)$.
A first step is the following theorem. 
\begin{theorem}\label{t7}
The ordinary BRST cohomology depending on the
fields $\phi^A$, the antifields $\phi^*_A$ and their derivatives is 
isomorphic to the cohomology of
$s_\omega$ depending on the fields
$\tilde\phi^A$ and their spatial derivatives~:
\begin{eqnarray}
H(s,[\phi^A,\phi^*_A])\simeq 
H(s_H,[\tilde\phi^A,\tilde\phi^*_A])\simeq 
H(s_\omega,[\tilde\phi^A\tilde]).
\end{eqnarray}
\end{theorem}
In other words, in a $s_H$ cocycle, 
one can get rid of the temporal derivatives and of the
antifields through the addition of a $s_H$ coboundary.
For a proof of this theorem, see Appendix C.
 
Starting from the bottom of the descent equations, one then proves
(see again Appendix C)
that a non trivial cocycle modulo $d$,
$a$, $s_H a+db=0$, given by
$a=\tilde a + dt a^0 $,
where $\tilde a$ does not involve the differential $dt$,
can be characterized by
\begin{eqnarray}
a=dt (-\{\phi^*_A\phi^A,\tilde b_0\}_{P,alt}+a^0_0)+\tilde a_0,\label{1}
\end{eqnarray}
verifying
\begin{eqnarray}
s_\omega\tilde a_0+\tilde d \tilde b_0=0\label{ch1}\\
s_\omega a^0_0 +\tilde d b^0_0-{\partial\over\partial t}\tilde
b_0+\{h,\tilde b_0\}_{P,alt}=0\label{ch}.
\end{eqnarray}
Here, $\tilde a_0,\tilde b_0,a^0_0$ and $b^0_0$ contain no antifields and no
time derivatives of the fields, while $\tilde b_0$ and $b^0_0$ satisfy 
analoguous 
equations to $\tilde a_0$ and $a^0_0$ for some $\tilde m_0, m^0_0$. 
In maximum form degree $n$, there is of course no $\tilde a$ and  
at the bottom, say  $n$, of the descent equations, 
$\tilde n_0$ and $n^0_0$ 
are $s_\omega$-cocycles\footnote{These equations have been first used in
\cite{Barnich} to compare anomalies in the Hamiltonian
and the Lagrangian formalism.}.
 
In the coboundary condition for such
cocycles $a=s_H c + de$, we have 
\begin{eqnarray}
c=dt(-\{\tilde\phi^*_A\tilde\phi^A,\tilde
e_0\}_{P,alt}+e^0_0)+\tilde c_0,\label{cob1}
\end{eqnarray}
giving the conditions
\begin{eqnarray}
\tilde a_0=s_\omega\tilde c_0+\tilde d\tilde e_0\label{sch}\\
a^0_0= -s_\omega c^0_0-\tilde d e^0_0+{\partial\over\partial t}\tilde
e_0-\{h,\tilde e_0\}_{P,alt}
\label{cob}
\end{eqnarray}
where $\tilde c_0,\tilde e_0,c^0_0$ and $e^0_0$ again contain 
no antifields and no
time derivatives of the fields, with analoguous equations holding for
$\tilde b_0,b^0_0$ in terms of $\tilde e_0, e^0_0,\tilde f_0,f^0_0$. 
In maximum form degree, there is no
$\tilde a, \tilde c$ and equation (\ref{sch}) is trivially satisfied.

In order to characterize the local BRST cohomology groups
$H^{g,k}(s_H|d)$, one can first find a basis for the vector space 
$H^{g,k}(s_\omega|\tilde d)$ in the space of antifield and time
derivative independent local forms with only spatial differentials
(most general non trivial solution for $\tilde a_0$). 
One then finds a basis for $H^{g+1,k-1}(s_\omega|\tilde d)$
(most general non trivial solution for $\tilde b_0$). 
One finally considers 
the subspace $l[H^{g+1,k-1}(s_\omega|\tilde d)]$ for which equation
(\ref{ch}) admits a particular solution $a^0_{0P}$. 
The general non trivial form
for $a^0_0$ is then given by $a^0_0=a^0_{0P}+\bar a^0_0$ where $\bar a^0_0$
belongs to $r[H^{g,k-1}(s_\omega|\tilde d)]$, which is the subspace of
$H^{g,k-1}(s_\omega|\tilde d)$ remaining non trivial under the more
general coboundary condition (\ref{cob}).

We thus get the following result on the relationship between the 
local BRST cohomology groups
in Lagrangian and Hamiltonian formalism :
\begin{theorem} \label{t6}
The local BRST cohomology groups are isomorphic to the
direct sum of the following three local cohomology groups of the
Hamiltonian formalism~:
\begin{eqnarray}
H^{g,k}(s|d)\simeq H^{g,k}(s_H|d)\simeq \nonumber\\
H^{g,k}(s_\omega|\tilde d)\oplus
l[H^{g+1,k-1}(s_\omega|\tilde
d)]\oplus r[H^{g,k-1}(s_\omega|\tilde d)].\label{iso}
\end{eqnarray}
\end{theorem}
Note that in maximal form degree $n$, the 
first group of the last line vanishes. This decomposition is in
general quite difficult to achieve in practice since it requires the
resolution of complicated equations. However, it corresponds to the
natural resolution of the spatio-temporal descent equations in the
Hamiltonian formalism and it is useful in principle, since it enables
one to relate the bracket and the antibracket.

{\bf Remark :}
The groups with prefix $r$ and $l$ appear also in the 
covariant analysis of the
descent equations in the following way. The descent equations provide
a homomorphism ${\cal D}~:H^{g,k}(s|d)\longrightarrow
H^{g+1,k-1}(s|d)$ with ${\cal D}[a]=[b]$ for $sa+db=0(\longrightarrow
sb+dc=0)$. The kernel of ${\cal D}$ can easily be shown to consist of
the vector space $H^{g,k}(s)$ seen as a subspace of $H^{g,k}(s|d)$,
i.e., the equivalence classes of $s$-cocycles with equivalence
relation determined by $s$ modulo $d$ exactness. We denote this kernel
by $r[H^{g,k}(s)]$.

The image of ${\cal D}$ is given by the classes $[b]\in
H^{g+1,k-1}(s|d)$, which can be lifted, i.e., such that there exists
$a$ with $sa+db=0$. We denote this space by $l[H^{g+1,k-1}(s|d)]$. 

This implies the isomorphism 
\begin{eqnarray}
H^{g,k}(s|d)\simeq
l[H^{g+1,k-1}(s|d)]\oplus r[H^{g,k}(s)]
\end{eqnarray}
and, by iteration,
\begin{eqnarray}
H^{g,k}(s|d)\simeq
\oplus_{i=0}^{k}l^i r[H^{g+i,k-i}(s)],
\end{eqnarray}
where in the last space ($i=k$) one can forget the $r$,
because there are no $d$ exact terms in form degree $0$. Note that
since 
$H^0(d)={\bf R}$, if $g=-k$, the last 
space has to be replaced by the space $\{e,
se=c, e\sim e +sf +c^\prime; c, c^\prime\in {\bf R}\}$ which is 
isomorphic to $H^0(s)/{\bf R}$.

In the Hamiltonian case above, we consider only the part of the
descent equations involving the exterior derivative with respect to
time~: $d^0=dt\ (d/dt)$.\qed

We now use theorem \ref{t6} to derive information on the antibracket
from the Poisson bracket induced in $H(s_\omega|\tilde d)$.
On the representatives of the local BRST cohomology groups determined
by equations (\ref{1})-(\ref{cob}), the local antibracket gives
\begin{eqnarray}
\{\hat a_1,\hat a_2\}=\{\phi^*_A\phi^A,
\{\hat{\tilde b}_1,\hat{\tilde b}_2\}_{P}\}_{P,alt}
-\{\hat a^0_1,\hat {\tilde b}_2\}_{P}
-(-)^{\varepsilon_{\hat {\tilde b}_1}}
\{\hat {\tilde b}_1, \hat a^0_2\}_{P}.
\end{eqnarray}
Hence, (i) the antibracket map can be entirely rewritten in terms of
the local Poisson bracket and
(ii) it is non trivial only if $l[H^{*,n-1}(s_\omega|\tilde d)]$ is
non trivial.

More precisely, according to the split of $H^{*,n}(s|d)$ in (\ref{iso})
to which corresponds the split of 
$a^0$ into $a^0_P$ and $\bar
a_0$, we see that the antibracket map (\ref{abm}) 
is completely determined by the
local Poisson bracket map induced in 
\begin{eqnarray}
\{\cdot,\cdot\}_P:l[H^{g_1+1,n-1}(s_\omega|\tilde
d)]\times l[H^{g_2+1,n-1}(s_\omega|\tilde d)]\nonumber\\
\longrightarrow
l[H^{g_1+g_2+2,n-1}(s_\omega|\tilde d)]
\end{eqnarray}
and by the local Poisson bracket map in 
\begin{eqnarray}
\{\cdot,\cdot\}_P:l[H^{g_1+1,n-1}(s_\omega|\tilde d)]
\times r[H^{g_2,n-1}(s_\omega|\tilde d)]\nonumber\\
\longrightarrow
r[H^{g_1+g_2+1,n-1}(s_\omega|\tilde d)].\label{2}
\end{eqnarray}
Hence the antibracket map is determined by the following 
matrix in maximum spatial form degree $n-1$ :
\begin{eqnarray}
\left( \begin{array}{cc}\{l[H^{g_1+1}(s_\omega|\tilde
d)], l[H^{g_2+1}(s_\omega|\tilde d)]\}_P & (-)^{\varepsilon_{g_1+1}}
\{l[H^{g_1+1}(s_\omega|\tilde
d)], r[H^{g_2}(s_\omega|\tilde d)] \}_P  \\
\{r[H^{g_1}(s_\omega|\tilde d)],l[H^{g_2+1}(s_\omega|\tilde
d)]\}_P & 0\end{array}\right).\label{3}
\end{eqnarray} 
Equations (\ref{2}) and (\ref{3}) mean 
in particular that $r[H^{*,n-1}(s_\omega|\tilde d)]$ is an abelian
subalgebra and an
ideal in the odd Lie algebra $(H^{*,n}(s|d),\{\cdot,\cdot\})$. 
We have thus proved :
\begin{theorem} The odd Lie algebra $(H^{*,n}(s|d),\{\cdot,\cdot\})$ 
is isomorphic to the semi-direct sum of the abelian Lie algebra
$r[H^{*,n-1}(s_\omega|\tilde d)]$ and the Lie algebra 
$(l[H^{*,n-1}(s_\omega|\tilde d)],\{\cdot,\cdot\}_P)$, where 
the action of $(l[H^{*,n-1}(s_\omega|\tilde d)]$ on 
$r[H^{*,n-1}(s_\omega|\tilde d)]$ is determined by the Poisson bracket map 
from one space to the other. By taking the quotient, 
the following isomorphism is seen to hold~: 
\begin{eqnarray}
(H^{*,n}(s|d)/r[H^{*,n-1}(s_\omega|\tilde d)]
,\{\cdot,\cdot\})\simeq(l[H^{*,n-1}(s_\omega|\tilde
d)],\{\cdot,\cdot\}_P).
\end{eqnarray}
\end{theorem}

The consequences of this result in the particular case of conserved currents,
i.e., for $g_1=g_2=-1, k=n$ are as follows.  
Using the results of \cite{BBH} in both the Lagrangian and 
the Hamiltonian context, the isomorphism (\ref{iso}) means 
that 

(i) the space of inequivalent Lagrangian conservation laws of the
first group is isomorphic to 
the subspace of spatial local functionals
in the coordinates and momenta, 
defined on the constraint surface $\tilde \Sigma$ and
gauge invariant on this surface, whose Poisson bracket with the first
class Hamiltonian $H_0$
plus the explicit time derivative vanishes on the 
contraint surface, 
\begin{eqnarray}
\{ Q=\int_{t=t_0} \tilde b_0 [\phi^i\pi_j\tilde ]\equiv\int_{t=t_0}
d^1x\dots d^{n-1}x \tilde j^0 [\phi^i\pi_j\tilde ],\nonumber\cr
[ Q, H_{0} ]_P+{\partial\over\partial t}Q=0|_{\tilde\Sigma},
Q\sim Q|_{\tilde\Sigma}\},
\end{eqnarray}
and 

(ii) the space of inequivalent Lagrangian conservation laws of the
second group is isomorphic to  a subspace of the characteristic
cohomology 
of the constraint surface in
spatial form degree $(n-1)-1$, the space of conservation laws
associated to the contraint surface, where two such conservation laws
have to be considered to be equivalent if they differ on the
constraint surface by a spatial superpotential 
and the total time derivative of a spatial current,
\begin{eqnarray}
\{\tilde j^k, \partial_k \tilde j^k = 0|_{\tilde \Sigma}, 
\tilde j^k \sim|_{\tilde \Sigma} \tilde j^k +\partial_j
\tilde S^{[jk]}+{\partial\over\partial t}\tilde f^k-
\{h_0,\tilde f^k\}_{P,alt}\}.
\end{eqnarray}

For example, the current corresponding to the Lagrangian current
$j^{\mu}=F^{0\mu}$ is given by the momenta $\pi^k$ in the case of
electromagnetism. 

The semi-direct sum structure holds also for the Lagrangian Dickey
algebra, but furthermore, we get from (\ref{3}) that (i) 
the algebra of
inequivalent conserved charges ${\cal Q}$ 
equipped with the induced Dickey bracket 
corresponds to the ordinary Poisson bracket algebra of conserved
inequivalent charges in the Hamiltonian
formalism and (ii) that the ideal $I$ of conserved currents without
charge forms an abelian subalgebra.

\section{Conclusion} 

We have shown what is the precise relationship  
between the antibracket map and various Lie algebras existing for
local gauge field theories. In the case of conserved currents, 
where ``covariant'' Poisson brackets are known, a direct 
comparision has been given. 

In the general case, the antibracket map is 
related to the Poisson bracket of the canonical formalism. 
The core of this analysis is the
relationship of the local BRST 
cohomologies in the Lagrangian and the Hamiltonian formalisms (i.e.,
the cohomologies modulo $d$ in the Lagrangian case and modulo $\tilde
d$ in the Hamiltonian one). This relationship turns
out to be somewhat more subtle than for the ordinary 
cohomologies, or the cohomologies modulo $\tilde d$, 
which are simply isomorphic. 

We have shown in particular what is 
the precise analog of the Lie algebra of 
inequivalent conserved currents in the Hamiltonian framework, which 
in turn allows some general statements on the structure of this Lie algebra
and could be useful for its actual computation.  

\section*{Acknowledgements}
The authors are grateful to Friedemann Brandt and 
Jim Stasheff for useful discussions. This
work has been supported in part by research contracts with the
F.N.R.S. and the Commission of the European Community.

\section*{Appendix A : 
Jet-spaces, variational bicomplex and Koszul-Tate resolution}

\renewcommand{\theequation}{A.\arabic{equation}}
\setcounter{equation}{0}

In this appendix, we recall briefly the construction of jet-bundles and 
of the variational bicomplex. We will construct a tricomplex containing
the horizontal, the vertical and the Koszul-Tate differentials.
The construction enhances the cohomological setup of the variational
bicomplex associated to possibly
degenerate partial differential equations by implementing the pullback
from the free bicomplex to the bicomplex of the surface 
defined by the equations through the homology of the
Koszul-Tate differential. (Different considerations on the
Batalin-Vilkovisky formalism in the context of the variational
bicomplex are given in \cite{McCloud}.)

Let us first recall some of the ingredients of the variational
bicomplex relevant for our purpose
(for a review see \cite{Olver,Saunders,Anderson} 
and the references to the original
literature therein). 
As we will not be 
concerned with global properties, we will work 
in local coordinates throughout. Consider a trivial fiber bundle 
\begin{eqnarray}
\pi : E={\bf M}^n \times F  \rightarrow {\bf M}^n
\end{eqnarray}
with local coordinates
\begin{eqnarray}
\pi : (x^\mu,\phi^i)\rightarrow (x^\mu)
\end{eqnarray}
where $\mu=0,\dots,n-1$ and
$i=1\dots, m$, with $F$ a manifold homeomorphic to ${\bf R}^m$
parametrized by the $\phi^i$.
For simplicity, we assume here that all
the $\phi^i$ are even, but all the considerations 
that follow could also be done in the case where the 
original bundle is a superbundle. 

The induced coordinates on the infinite jet bundle 
\begin{eqnarray}
\pi^\infty : J^\infty(E)={\bf M}^n\times F^\infty 
\rightarrow  {\bf M}^n
\end{eqnarray}
of jets
of sections on ${\bf M}^n$ are given by 
\begin{eqnarray}
(x^\mu,\phi^i,\phi^i_{\mu},
\phi^i_{\mu_1\mu_2},\dots,).
\end{eqnarray}
Let $\Omega^p(J^\infty(E))$ be the 
local differential forms on $J^\infty(E)$. The exterior differential $d_T$ 
is 
split into horizontal and vertical differentials: $d_T=d_H+d_V$, with
\begin{eqnarray}
d_H=dx^\mu \partial_\mu,\quad \partial_\mu={\partial^L\over\partial x^\mu}+
\phi^i_{(\nu)\mu}{\partial^L\over\partial \phi^i_{(\nu)}}
\end{eqnarray}
and 
\begin{eqnarray}
d_V\phi^i_{(\nu)}=d\phi^i_{(\nu)}-dx^\mu\phi^i_{(\nu)\mu},\  d_V x^\mu=0.
\end{eqnarray}

Note that everywhere else in the paper, we have omitted the
subscript $H$ on the horizontal differential and that we have
introduced the more compact notation $\phi^i_{(\nu)}\equiv
\partial_{(\nu)}\phi^i$ for the independent coordinates corresponding
to the derivatives of the fields.

Furthermore, we have
\begin{eqnarray}
d_Hd_V+d_Vd_H=d_H^2=d_V^2=0. 
\end{eqnarray}
A local $p$-form of 
$\Omega^p(J^\infty(E))$ can then be written as a sum of terms of the form 
$f[\phi]dx^{\mu_1}\dots dx^{\mu_r}d_V\phi^{i_1}_{(\nu_1)}
\dots d_V\phi^{i_s}_{(\nu_s)}$
of horizontal degree $r$ and vertical degree $s$
with $r+s=p$ and $f[\phi]$ a smooth functions of $x^\mu$, $\phi^i$
and a finite number of their derivatives. The free variational bicomplex is
the double complex $(\Omega^{*,*}((J^\infty(E)),d_H,d_V)$ of differential 
forms on $(J^\infty(E))$. 

\begin{eqnarray}
\hspace*{-1.8cm}\begin{array}{ccccccccccccc}
 & & & & \vdots & &
& &\vdots & & \vdots \\
 & & & & d_V\uparrow & & & & d_V\uparrow & &
\delta_V\uparrow \\
 & & 0 & \longrightarrow
&
\Omega^{0,2}(J^\infty(E)) &
\stackrel{d}{\longrightarrow} & \dots &
\stackrel{d}{\longrightarrow}
 & \Omega^{n,2}(J^\infty(E)) & \stackrel{\int}\longrightarrow & {\cal
F}^2(J^\infty(E)) &
\longrightarrow & 0 \\
 & & & & d_V\uparrow & & & & d_V\uparrow & &
\delta_V\uparrow \\

 & & 0 & \longrightarrow
&
\Omega^{0,1}(J^\infty(E)) &
\stackrel{d}{\longrightarrow} & \dots &
\stackrel{d}{\longrightarrow}
 & \Omega^{n,1}(J^\infty(E)) & \stackrel{\int}\longrightarrow
 & {\cal F}^1(J^\infty(E)) &
\longrightarrow & 0 \\
 & & & & d_V\uparrow & & & & d_V\uparrow & &
\delta_V\uparrow \\
0 & \longrightarrow & {\bf R} &
\longrightarrow &
\Omega^{0,0}(J^\infty(E)) & \stackrel{d}{\longrightarrow}
& \dots  & \stackrel{d}{\longrightarrow} &
\Omega^{n,0}(J^\infty(E)) &
 \stackrel{\int}\longrightarrow & {\cal F}^0(J^\infty(E)) 
& \longrightarrow & 0 \\
 
 & & & & \uparrow & & & & \uparrow & &
\uparrow \\

0 & \longrightarrow & {\bf R} &
\longrightarrow &
\Lambda^0({\bf M^n}) & \stackrel{d}{\longrightarrow}
& \dots  & \stackrel{d}{\longrightarrow} &
\Lambda^n({\bf M^n}) &
\longrightarrow & 0 &  &  \\
& & & & \uparrow & & & &  \uparrow & &   \\
& & & & 0 & & & &  0 & &   \\
\end{array}\nonumber
\end{eqnarray}

The important property of this bicomplex is 
that all the rows and columns of the above diagram are exact
\cite{Olver,Saunders,Dickey}.
The integral sign $\int$  denotes the 
projection, for each vertical degree $s$, of horizontal $n$-forms onto
the space of local functional forms 
${\cal F}^s$, i.e., the space of equivalence 
classes obtained by identifying exact 
horizontal $n$-forms with zero: ${\cal F}^s=\Omega^{n,s}/d_H\Omega^{n-1,s}$.
$\delta_V$ is the induced action of the vertical derivative in ${\cal F}^s$:
$\delta_V \int\omega^{n,s}=\int d_V\omega^{n,s}$.
An evolutionary vector field on $E$ is given by 
$v_Q=Q^i[\phi]{\partial^L\over\partial \phi^i}$. Its prolongation is given
by $\delta_Q=\partial_{(\nu)}Q^i{\partial^L\over\partial \phi^i_{(\nu)}}$.
Because $[\delta_Q,\partial_\mu]=0$, 
the contraction of a functional form with 
the prolongation of evolutionary vector fields is well defined.

A system ${\cal R}$ of $k$-th order partial differential equations on 
$E$,
\begin{eqnarray}
{\cal R}_a(x^\mu,\phi^i,\phi^i_{\mu},\dots,\phi^i_{\mu_1\dots\mu_k})=0
\qquad a=1,\dots, l,
\end{eqnarray}
defines a subbundle ${\cal R}\rightarrow {\bf R}^n$ of $J^k(E)\rightarrow
{\bf R}^n$.
We shall assume that the equations ${\cal R}_a=0,
\partial_\mu{\cal R}_a=0,\dots,
\partial_{\mu_1\dots\mu_s}{\cal R}_a=0$ 
define, for each $x^\mu$, a smooth surface and provide a 
regular representation of this surface in the vector spaces 
$F^{s+k}$ for each $s$,
i.e., the equations can be split into independent equations ($L_m$)
which can be locally
taken as first coordinates in a new, regular, coordinate system in the
vicinity of the surface defined by the equations, and into dependent
ones ($L_\Delta$) which hold as a consequence of the independent ones. 

This implies that one can split locally the $\phi^i$ and their 
derivatives 
up to order $s+k$ into independent variables $x_A$ not constrained by the 
equations and dependent variables $z_\alpha$ in such a way that the 
equations ${\cal R}_a=0,\dots,\partial_{\mu_1\dots\mu_s}{\cal R}_a=0$
are equivalent to $z_\alpha =z_\alpha(x_A,
L_m)$. A local coordinate system adapted to the equations 
is then given
by $(x^\mu,x_A,L_m)$ in $J^{s+k}(E)$. How this works in detail for
Yang-Mills
theory, gravity or two-form fields, is discussed in \cite{BBH}.
The infinite prolongation ${\cal R}^\infty$ of ${\cal R}$, i.e., 
the given sets
of equations and all their total derivatives, 
defines a subbundle in
$J^\infty(E)$. In the sequel, by ``stationary
surface" or by ``on-shell"
we mean that we are on the subbundle defined by ${\cal R}_a=0$ and an 
appropriate number of its derivatives, depending on the space $J^l(E)$
under consideration.

A consequence of the regularity condition is that any function $f[\phi]$
which vanishes on the stationary surface, $f\approx 0$, 
can be written as a combination of 
the equations defining this surface \cite{Olver,Henneaux}. 

The
knowledge of the split of the equations into dependent and independent
ones allows one to find a locally complete set of non trivial 
local reducibility operators in $J^l(E)$, i.e., operators 
$R^{+a(\mu)}_{a_1}\partial_{(\mu)}$ for some local functions
$R^{+a(\mu)}_{a_1}[\phi]$ on $J^l(E)$ which do not all vanish on-shell
such that
\begin{eqnarray}
R^{+a(\mu)}\partial_{(\mu)}{\cal R}_a=0
\end{eqnarray}
and verifying the property that 
if $\lambda^{+a(\mu)}\partial_{(\mu)}{\cal R}_a=0$ for some
local functions $\lambda^{+a(\mu)}[\phi]$ on $J^l(E)$, then
\begin{eqnarray}
\lambda^{+a(\rho)}\partial_{(\rho)}=\lambda^{+a_1(\lambda)}
\partial_{(\lambda)}( R^{+a(\mu)}_{a_1}\partial_{(\mu)}\cdot)+
\mu^{a(\mu)b(\nu)}(\partial_{(\nu)} {\cal R}_b)\partial_{(\mu)}
\label{A10}
\end{eqnarray} for
some local functions $\lambda^{+a_1(\lambda)}[\phi]$, and 
$\mu^{a(\mu)b(\nu)}$
on $J^l(E)$,
where $\mu^{a(\mu)b(\nu)}=-
\mu^{b(\nu)a(\mu)}$. Furthermore, the first term of the right hand
side of equation (\ref{A10}) can be assumed to be absent if the
functions  $\lambda^{+a(\mu)}$ vanish on-shell \cite{Henneaux}. Such
reducibility operators will be called trivial because they exist for
any gauge theory. 

For simplicity, we will assume here that the reducibility operators
are themselves irreducible in the sense that if 
$\lambda^{+a_1(\lambda)}R^{+a(\mu)}\partial_{(\mu)}$ vanishes on the
stationary surface, the functions
$\lambda^{+a_1(\lambda)}$ vanish on the stationary surface.
All the considerations that follow can be generalized to the case
with higher order reducibility operators at the price of increasing the
number of additional generators introduced below like in
\cite{Henneaux}.  

The variational bicomplex $(\Omega^{*,*}({\cal R}^\infty), d_H, d_V)$
of the differential equations ${\cal R}$ is the 
pull-back of the variational bicomplex from $J^\infty(E)$ to 
${\cal R^\infty}$. With the previous assumptions, it is straightforward to
verify that $\Omega^{*,*}({\cal R}^\infty)$ is locally isomorphic to the 
forms in $dx^\mu$ and $d_V x_A$ with coefficients that are smooth functions
in the $x^\mu,x_A$. The columns of this bicomplex remain exact, because
the contracting homotopy \cite{Olver} which allows to prove exactness
in the free case still holds when we consider only $d_Vx_A$'s. 
There exist however non trivial cohomology
groups along the rows. 

The Koszul-Tate resolution of this bicomplex is obtained by a 
straightforward generalization of the Koszul-Tate resolution of the 
stationary surface ${\cal R}^\infty$ \cite{Henneaux}.
One considers the superbundle 
\begin{eqnarray}
\pi: K={\bf R}\times(F\oplus\Phi^*\oplus C^*) 
\longrightarrow {\bf R}
\end{eqnarray}
and the associated free variational bicomplex
$(\Omega^{*,*}(J^\infty(K)), d_H, d_V)$. $\Phi^*$ is the vector space 
with coordinates  
the Grassmann odd $\phi^*_a$
and is of dimension $l$, the number of original equations. $C^*$ the
vector space with coordinates the Grassmann even 
$C^*_{a_1}$ and its dimension equals the number of 
non-trivial reducibility operators $R^{+a(\mu)}_{a_1}\partial_{(\mu)}$.
The Koszul-Tate differential $\delta$ is defined on 
$\Omega^{*,*}(J^\infty(K))$ by 
\begin{eqnarray}
\delta x^\mu=\delta dx^\mu=\delta\phi^i=0,\nonumber\\
\delta \phi^*_a = {\cal R}_a,\quad \delta C^*_{a_1}=R^{+a(\mu)}_{a_1}
\partial_{(\mu)}\phi^*_a \nonumber\\
\delta d_H+d_H \delta=0=\delta d_V+d_V\delta
\end{eqnarray}
and is extended as a left antiderivation.
The associated grading is obtained from the eigenvalues of the
antighost number operator 
defined by
\begin{eqnarray}
antigh=\phi^*_{a(\mu)}{\partial^L\over\partial\phi^*_{a(\mu)}}+
d_V \phi^*_{a(\mu)}{\partial^L\over\partial d_V\phi^*_{a(\mu)}}
\nonumber\\+
2 C^*_{a_1(\mu)}{\partial^L\over\partial C^*_{a_1(\mu)}}+
2d_V C^*_{a_1(\mu)}{\partial^L\over\partial d_VC^*_{a_1(\mu)}}.
\end{eqnarray}
As in \cite{Henneaux}, one can then prove that 
\begin{eqnarray}
H_0(\delta,\Omega^{*,*}_0(J^\infty(K)))\simeq\Omega^{*,*}(J^\infty(E))
/{\cal N}\simeq \Omega^{*,*}({\cal R}^\infty)
\end{eqnarray}
and that 
\begin{eqnarray}
H_k(\delta,
\Omega^{*,*}_k(J^\infty(K)))=0,\quad for\ k>0.
\end{eqnarray}
Here, ${\cal N}$ is the ideal of forms such that each term contains at least
one of the terms  
$\partial_{(\mu)}{\cal R}_a$ or  
$d_V \partial_{(\mu)}{\cal R}_a$. Hence, locally, 
the quotient is isomorphic to the forms in 
$dx^\mu$ and $d_V x_A$ with coefficients that are smooth functions
in the $x^\mu,x_A$. By using a partition of unity, we then get 
the last 
isomorphism in the above equation.
This means that the diagram 
\begin{eqnarray}
\dots\stackrel{\delta}{\longrightarrow}\Omega_k^{*,*}(J^\infty(K))
\stackrel{\delta}{\longrightarrow}\Omega_{k-1}^{*,*}(J^\infty(K))
\stackrel{\delta}{\longrightarrow}\dots\nonumber\\
\dots
\stackrel{\delta}{\longrightarrow}\Omega_1^{*,*}(J^\infty(K))
\stackrel{\delta}{\longrightarrow}
\Omega^{*,*}({\cal R}^\infty)\longrightarrow 0\nonumber
\end{eqnarray}
is exact.

In the three dimensional grid corresponding to the tricomplex 
\begin{eqnarray}
(\Omega^{*,*}_*(J^\infty(K), d_H, d_V,\delta))
\end{eqnarray}
augmented by the 
projection on local functionals in the $d_H$ direction and  
by the projection on the bicomplex for the partial differential equations
$(\Omega^{*,*}({\cal R}^\infty), d_H, d_V)$ in the $\delta$ direction,
except for the rows of this last
complex, the sequences are exact in all directions.

The advantage of this cohomological resolution of the 
variational bicomplex for partial differential equations is 
that the non trivial
cohomology groups $H^{r,*}(d_H,\Omega^{r,*}({\cal R}^\infty))$ are 
given by relative cohomology groups in the free tricomplex
\begin{eqnarray}
H^{r,*}(d_H,\Omega^{r,*}({\cal R}^\infty))\simeq
H^{r,*}_0(d_H|\delta,\Omega^{r,*}_0(J^\infty(K)))\label{iso2}.
\end{eqnarray}
Since $H^{q,*}_*(d_H,\Omega^{q,*}_*(J^\infty(K)))=0$ for $0<q<n$ and

\noindent $H^{*,*}_k(\delta,\Omega^{*,*}_k(J^\infty(K)))=0$ 
for $k>0$, one can for instance apply 
the method of diagram chasing (or ``snake lemma") in the horizontal
and $\delta$ directions to get, for $(r,s)\neq(0,0)$,
\begin{eqnarray}
H^{r,s}_0(d_H|\delta,\Omega^{r,s}_0(J^\infty(K)))\simeq
H^{r+1,s}_1(d_H|\delta,\Omega^{r,s}_1(J^\infty(K)))\simeq\dots
\nonumber\\ \simeq
H^{n-1,s}_{n-r-1}(d_H|\delta,\Omega^{n-1,s}_{n-r-1}(J^\infty(K)))
\label{iso3}.
\end{eqnarray}
For $(r,s)=(0,0)$, the same chain of isomorphisms remain true if one 
replaces the 
first element in the chain by 
$H^{0,0}(d_H|\delta,\Omega^{0,0}_0(J^\infty(K)))/{\bf R}$.
Furthermore, like in \cite{DuboisViolette}, one proves that:
\begin{eqnarray}
{H^{r,*}_k(d_H|\delta,\Omega^{r,*}_k(J^\infty(K)))\over
p^{\#}H^{r,*}_k(d_H,\Omega^{r,*}_k(J^\infty(K)))}\simeq
{H^{r+1,*}_{k+1}(\delta|d_H,\Omega^{r+1,*}_{k+1}(J^\infty(K)))\over
p^{\#}H^{r+1,*}_{k+1}(\delta,\Omega^{r+1,*}_{k+1}(J^\infty(K)))}
\end{eqnarray}
where $p^{\#}$ denotes the natural inclusion of an absolute 
cohomology group as a relative cohomology group.
Using the results on the cohomology of $d_H$ and $\delta$, 
these relations reduce to 
\begin{eqnarray}
H^{0,0}_0(d_H|\delta,\Omega^{0,0}_0(J^\infty(K)))/{\bf R}\simeq
H^{1,1}_1(\delta |d_H,\Omega^{1,1}_1(J^\infty(K)))\\
H^{r,*}_k(d_H|\delta,\Omega^{r,*}_k(J^\infty(K)))\simeq
H^{r+1,*}_{k+1}(\delta |d_H,\Omega^{r+1,*}_{k+1}(J^\infty(K))),
\nonumber\\
(r,s,k)\neq (0,0,0), r<n.\label{iso4}
\end{eqnarray}

\section*{Appendix B : Local brackets and surface terms}

\renewcommand{\theequation}{B.\arabic{equation}}
\setcounter{equation}{0}

In the first part of this appendix, we want to calculate explicitly
the total divergences that arise in the Jacobi identity for the local
(anti)bracket. 

Let $z^a=(\phi^A;\phi^*_A)$ and $\zeta^{ab}= \left( \begin{array}{cc}0
&{\delta_B}^A\\-{\delta_B}^A&0\end{array}\right)$. Let
$\tilde{(\nu)}_\mu$ denote the number of times the index $\mu$ appears
in the multiindex $(\nu)$. The higher Euler operators \cite{Olver}
are uniquely defined by the expression 
\begin{eqnarray}
\delta_Q f = \partial_{(\nu)}(Q^a{\delta^L f\over\delta z^a_{(\nu)}}).
\end{eqnarray}
Let us furthermore define the ``generalized Hamiltonian vector field''~:
\begin{eqnarray}
\bar a^b=({\delta^R\hat a \over\delta z^{a}})\zeta^{ab}.
\end{eqnarray} 
Then the local
antibracket in the space of integrands (\ref{la}) can be rewritten as 
\begin{eqnarray}
\{a_1,a_2\}=d^nx [\partial_{(\mu)}({\delta^R\hat a_1 \over\delta z^{a}})
\zeta^{ab}{\partial^L\hat a_2 \over\partial z^{b}_{(\mu)}})
\nonumber\\-{\tilde{(\nu)}_\mu+1\over|\nu|+1}\partial_{\mu (\nu)}
({\delta^R\hat a_1\over\delta z^{a}}\zeta^{ab}
{\delta \hat a_2\over \delta z^{b}_{\mu(\nu)}})]
\equiv \delta_{\bar a_1}a_2-d I_{\bar a_1} a_2.\label{A1}
\end{eqnarray}
This expression implies that the graded Leibnitz rule 
holds up to a total divergence. 

We have pointed out in the text that the local antibracket (\ref{A1})
does not satisfy the graded 
Jacobi identity strictly, but only up to a total
divergence. Similarily, in the Hamiltonian theory, the Poisson bracket
among local functions of the fields, their conjugate momenta and their
derivatives, 
\begin{eqnarray}
\{\hat a_1,\hat a_2\}_P={\tilde\delta^R\hat a_1\over\delta\phi^i}\sigma^{ij}
{\tilde\delta^L\hat a_2\over\delta\phi^j},\label{B5}
\end{eqnarray}
satisfies the Jacobi
identity $\{\hat a,\{\hat b,\hat c\}\} + cyclic=0$ only up to a
(spatial) total divergence. In equation (\ref{B5}),  
$\phi^i$ collectively denotes the fields and their conjugate
momenta, the tilde superscript denotes spatial Euler-Lagrange
derivatives and $\sigma^{ij}= \left( \begin{array}{cc}0
&{\delta_j}^i\\-{\delta_j}^i&0\end{array}\right)$.

For definitness, we shall evaluate here explicitly the boundary terms 
in the Jacobi identity in the Hamiltonian case and assume that 
the fields $\phi^i$ and the densities $\hat a,\hat b,$ and $\hat c$
are all even. We will however not write explicitly the tilded
superscript to indicate the spatial derivatives.
The calculation for the local antibracket or the local extended
Poisson bracket
is simply a matter of taking care of the sign factors.

We will need the following lemma~:
\begin{eqnarray}
(-\partial)_{(\alpha)}
(f{\partial\over\partial\phi^i_{(\alpha)}}\partial_\beta
g )= - (-\partial)_{(\alpha)}
(\partial_\beta f{\partial\over\partial\phi^i_{(\alpha)}}
g ).\label{lem}
\end{eqnarray}
The proof of this lemma follows from a straightforward extension
of the proof of
${\delta\over\delta\phi^i}(\partial_\beta g)=0$ in \cite{DeDonder}. 

A direct calculation, using the analog of (\ref{A1}) for the Poisson
bracket and the fact that the Euler-Lagrange derivatives annihilate
total divergences, yields
\begin{eqnarray}
\{a,\{b,c\}\}+cyclic=\{a,\{b,c\}\}-\{b,\{a,c\}\}-\{\{a,b\},c\}
\label{Leibnitz}\\
=\delta_{\bar a}\delta_{\bar b}c - \delta_{\bar b}\delta_{\bar a}c 
-\delta_{\bar{\{a,b\}}}c 
\nonumber\\-dI_{\bar a}(\delta_{\bar b}c)+
dI_{\bar b}(\delta_{\bar a}c)
+dI_{\bar{\{a,b\}}}c
\end{eqnarray}
We have
\begin{eqnarray}
(\delta_{\bar a}\delta_{\bar b}-\delta_{\bar b}\delta_{\bar a})c=
\delta_{\bar d}c
\end{eqnarray}
with
\begin{eqnarray}
\bar d^i\equiv(\delta_{\bar a}({\delta \hat b\over\delta
\phi^j})\sigma^{ji}-(\hat a\leftrightarrow\hat b)
\nonumber\\
={\delta\over\delta \phi^j}(\delta_{\bar a}\hat b)\sigma^{ji}-
(-\partial)_{(\alpha})[{\partial\over\partial\phi^j_{(\alpha)}}
(\partial_{(\beta)}{\delta \hat a\over\delta\phi^k})
\sigma^{kl}{\partial\hat b\over\partial \phi^l_{(\beta)}}]\sigma^{ji}
-(\hat a\leftrightarrow\hat b)
\nonumber\\
={\delta\over\delta
\phi^j}\{\hat a,\hat b\}\sigma^{ji}-
(-\partial)_{(\alpha})[{\partial\over\partial\phi^j_{(\alpha)}}
({\delta \hat a\over\delta\phi^k})
\sigma^{kl}{\delta\hat b\over\delta\phi^l}]\sigma^{ji}
-(\hat a\leftrightarrow\hat b)
\nonumber\\
=2{\delta\over\delta
\phi^j}\{\hat a,\hat b\}\sigma^{ji}-
(-\partial)_{(\alpha})[{\partial\over\partial\phi^j_{(\alpha)}}
({\delta \hat a\over\delta\phi^k}
\sigma^{kl}{\delta\hat b\over\delta\phi^l})]\sigma^{ji}
\nonumber\\
={\delta\over\delta
\phi^j}\{\hat a,\hat b\}\sigma^{ji}=\bar{\{a,b\}}^i
\end{eqnarray}
To get the line before last, we have used repeatedly the abovementioned
lemma (\ref{lem}). Hence,
\begin{eqnarray}
\{a,\{b,c\}\}+cyclic=d(-I_{\bar a}(\delta_{\bar b}c)+
I_{\bar b}(\delta_{\bar a}c)
+I_{\bar{\{a,b\}}}c).
\end{eqnarray}
This is the desired formula.\footnote{It also follows from this proof
that the alternative bracket given by $\{a,b\}_{alt}=\delta_{\bar a}b$
satisfies a strict Jacobi identity under Leibnitz form (defined by the
right hand side of (\ref{Leibnitz})), using furthermore the fact that
$\bar{\{a,b\}}_{alt}=\bar{\{a,b\}}$.}\qed

We now prove that the expressions in
equation (\ref{br1}) for the
Dickey bracket are equivalent to the definition in equation
(\ref{x}). 
Let us write terms which vanish on-shell by $\delta()$. By applying
the lemma (\ref{lem}), we find that, if $X$ is the characteristic 
of a variational symmetry, the following equation holds~: 
\begin{eqnarray}
\delta_X ({\delta\hat{{\cal L}}_0\over\delta\phi^i})Y^id^nx=
-(-\partial)_{(\mu)}[{\partial \over\partial
\phi^i_{(\mu)}}(\partial_{(\nu)}X^j )
{\partial \hat{{\cal L}}_0\over\partial\phi^j_{(\nu)}}]Y^id^nx=\nonumber\\
-(-\partial)_{(\mu)}[{\partial \over\partial
\phi^i_{(\mu)}}(X^j){\delta\hat{{\cal L}}_0\over\delta\phi^j}]Y^id^nx
=-\delta_Y(X^j){\delta\hat{{\cal L}}_0\over\delta\phi^j}d^nx
+d\delta()\label{meqt}
\end{eqnarray}
Let us evaluate $d(-\delta_{X_1}j_2)$. Using (\ref{meqt}) twice, we get,
\begin{eqnarray}
d(-\delta_{X_1}j_2)=-\delta_{X_1}({\delta\hat{{\cal
L}}_0\over\delta\phi^i}X^i_2)d^nx+d\delta()=\nonumber\\
-\delta_{X_1}(X^i_2)
{\delta\hat{{\cal L}}_0\over\delta\phi^i}d^nx+\delta_{X_2}(X^j_1)
{\delta\hat{{\cal L}}_0\over\delta\phi^i}d^nx+d\delta()\label{br2}\\
=d(\delta_{X_2}j_1)+d\delta().
\end{eqnarray}
{}From this equation it also follows immediately that 
\begin{eqnarray}
d(-\delta_{X_1}j_2)={1\over
2}d(\delta_{X_2}j_1)-d(\delta_{X_1}j_2)+d\delta()\label{br3}. 
\end{eqnarray}
Using the triviality of the cohomology of $d$ in form degree $n-1$
($>0$) implies the first two expressions in equation (\ref{br1}). 

{}From equation (\ref{br2}), it follows that 
\begin{eqnarray}
d(-\delta_{X_1}j_2)=[\delta_{[X_1,X_2]}\hat{{\cal L}}_0\nonumber\\
-{\tilde{(\nu)}_\mu+1\over|\nu|+1}\partial_{\mu(\nu)}({\delta
\hat{{\cal L}}_0\over\delta
\phi^i_{(\nu)\mu}}[X_1,X_2^i)]d^nx+d\delta ().
\end{eqnarray}
But we also have 
\begin{eqnarray}
\delta_{[X_1,X_2]}\hat{{\cal L}}_0
=(\delta_{X_2}\delta_{X_1}-\delta_{X_1}\delta_{X_2})
\hat{{\cal L}}_0\nonumber\\=\partial_\mu
(\delta_{X_2}j_1^\mu-\delta_{X_1}j_2^\mu)+\nonumber\\
{\tilde{(\nu)}_\mu+1\over|\nu|+1}\partial_{\mu(\nu)}[\delta_{X_2}
({\delta \hat{{\cal L}}_0\over\delta\phi^i_{(\nu)\mu}}
X_1^{i})-\delta_{X_1}
({\delta \hat{{\cal L}}_0\over\delta\phi^i_{(\nu)\mu}}
X_2^{i})]
\end{eqnarray}
This implies
\begin{eqnarray}
d(-\delta_{X_1}j_2)=d(\delta_{X_2}j_1
-\delta_{X_1}j_2)\nonumber\\+
{\tilde{(\nu)}_\mu+1\over|\nu|+1}\partial_{\mu(\nu)}
(\delta_{X_2}({\delta \hat{{\cal L}}_0\over\delta
\phi^i_{(\nu)\mu}})X_1^{i}-\delta_{X_1}
({\delta \hat{{\cal L}}_0\over\delta\phi^i_{(\nu)\mu}})
X_2^{i})d^nx+d\delta ().
\end{eqnarray}
Using (\ref{br3}), we find the last expression of (\ref{br1})~:
\begin{eqnarray}
d(-\delta_{X_1}j_2)=-{\tilde{(\nu)}_\mu+1\over|\nu|+1}
\partial_{\mu(\nu)}
(\delta_{X_2}({\delta \hat{{\cal L}}_0\over\delta\phi^i_{(\nu)\mu}})
X_1^{i}\nonumber\\-\delta_{X_1}({\delta \hat{{\cal L}}_0
\over\delta\phi^i_{(\nu)\mu}})X_2^{i})d^nx+d\delta()\label{krol}.\qed 
\end{eqnarray}

In the last part of the appendix, we establish the relationship
between the antibracket map and the Dickey bracket.
As explained before theorem \ref{t2}, we have to evaluate
$\delta\{a_1,a_2\}$, where $a=d^nx \phi^*_i X^i$ with $X^i$ defining a
variational symmetry~:
\begin{eqnarray}
\delta \{a_1,a_2\}=(d^nx {\delta X^i_1\over\delta\phi^j}X^j_2-
{\delta X^i_2\over\delta\phi^j}X^j_1)
{\delta{\cal L}_0\over\delta\phi^i}\nonumber\\
=d^nx(\delta_{X_2}X^i_1-\delta_{X_1}X^i_2)
{\delta{\cal L}_0\over\delta\phi^i} +d\delta ()
=d(-\delta_{X_1}j_2)+d\delta(),
\end{eqnarray} 
where we have used (\ref{br2}) in order to
get the last equality. This proves that to the antibracket map of two
classes in $H^n_1(\delta|d)$ corresponds the Dickey bracket of the
corresponding currents.\qed

\section*{Appendix C : Descent equations in the Hamiltonian
formalism} 

\renewcommand{\theequation}{C.\arabic{equation}}
\setcounter{equation}{0}

We analyze in this appendix first of all the relationship between the
cohomology of $s_H$ defined in equation (\ref{sh}) and the cohomology of
$s_\omega$, thereby proving theorem \ref{t7}. 
Then we analyze the spatio-temporal descent equations of $s_H$ by
choosing representatives appropriate to the Hamiltonian formalism,
proving equations (\ref{1})-(\ref{cob}).

\noindent {\bf Cohomology of $s_H$ and $s_\omega$}

The cocyle $n$ in $s_H n=0$ depends on the coordinates
$x^\mu,\partial_{(\mu)}\tilde\phi^A,\partial_{(\mu)}\tilde\phi^*_A$.
Consider the change of coordinates which consists in replacing the time
derivatives of the fields and all their derivatives by the
$\partial_{(\mu)}{\cal L}_A$. In the new coordinates $n$ depends on 
$x^\mu,\partial_{(k)}\tilde\phi^A,\partial_{(\mu)}\tilde\phi^*_A,
\partial_{(\mu)}{\cal L}_A$.
Using $s_\omega \Omega=s_\omega H=0$ and the identitiy
\begin{eqnarray}
-s_\omega\{\tilde\phi^*_A\tilde\phi^A,\cdot\}_{P,alt}+
\partial_{(k)}{\tilde\delta^R\over\delta\tilde\phi^C}(
\sigma^{CA}\{\tilde\phi^*_B\tilde\phi^B,\omega\}_{P,alt})
{\partial^L\over\partial(\partial_{(k)}\tilde\phi^A)}\nonumber\\
=
\{\tilde\phi^*_A\tilde\phi^A,s_\omega\cdot\}_{P,alt}\label{com},
\end{eqnarray}
we find that $s_H {\cal L}_A=0$. This means 
that in the new coordinate system 
\begin{eqnarray}
s_H=s_\omega+\partial_{(\mu)}{\cal L}_A
{\partial^L\over\partial(\partial_{(\mu)}\tilde\phi^*_A)},
\end{eqnarray}
where $s_\omega$ is restricted to spatial derivatives. 
Introducing the contracting homotopy
\begin{eqnarray}
\rho= \partial_{(\mu)}\tilde\phi^*_A
{\partial^L\over\partial(\partial_{(\mu)}{\cal L}_A)},
\end{eqnarray}
the anticommutator $\{s_H,\rho\}=N=z^\alpha{\partial^L\over\partial
z^\alpha}$ is the operator counting the number
of coordinates $z^\alpha\equiv\partial_{(\mu)}\tilde\phi^*_A,
\partial_{(\mu)}{\cal L}_A$. The standard argument is then that 
\begin{eqnarray}
n=n(z^\alpha=0)+\int_0^1 {d\lambda\over\lambda}(N n)[\lambda
z^\alpha]\label{hf} \\
=n_0+s_H (\int_0^1 {d\lambda\over\lambda}(\rho n)[\lambda
z^\alpha].
\end{eqnarray}

The cocycle condition now reduces to $s_\omega n_0 =0$ and the coboundary
condition $n_0=s_H p$ reduces to $n_0=s_\omega p_0$. Indeed, applying
$N$ to the coboundary condition implies that $Ns_H p=0$. Using
\begin{eqnarray}
[N,s_H]=0
\end{eqnarray}
and the same decomposition of $p$ as for $n$ 
in (\ref{hf}), this
equation implies that $s_Hb=s_\omega b_0$.

This proves theorem \ref{t7}.\qed

In order to analyze the spatio-temporal descent equations for $s_H$,
we start form the bottom, which we can assume to be of the form $n_0$
as above. We then want to know under what conditions $n_0$ can be
lifted, i.e.,
what are the conditions for the existence of $m$ such that
$sm+dn_0=0$. We will now prove in particular 
the crucial lemma that $m$ can be
assumed to be independent of the coordinates $\partial_{(\mu)}{\cal
L}_A$, with a linear dependence in the antifields   
$\partial_{(k)}\tilde\phi^*_A$ only in the terms involving the
differential $dt$.

\noindent {\bf First lift from the bottom of the descent equations}

The spatial exterior differential has the same form in
the new coordinate system as it had in the old one. The total time
derivative, however, is given by
\begin{eqnarray}
{d\over dt}={\partial\over\partial t}+\partial_{(k)l+1}\tilde\phi^*_A
{\partial^L\over\partial(\partial_{(k)l}\tilde\phi^*_A)}+
\partial_{(k)l+1}{\cal L}_A
{\partial^L\over\partial(\partial_{(k)l}{\cal L}_A)}+\nonumber\\
\partial_{(k)}\sigma^{CA}[-{\cal
L}_C-{\tilde\delta^R 
h\over\delta\tilde\phi^C}
-{\tilde\delta^R\over\delta\tilde\phi^C}
\{\tilde\phi^*_B\tilde\phi^B,\omega\}_{P,alt}]
{\partial^L\over\partial(\partial_{(k)}\tilde\phi^A)}. 
\end{eqnarray}

We then decompose $m$ and $n_0$ into pieces respectively containing 
the differential $dt$ or not
($m^0,n^0_0$ and $\tilde m,\tilde n_0$). The cocycle condition
splits into~:
\begin{eqnarray}
s_H\tilde m+\tilde d\tilde n_0=0,\ s_H m^0+\tilde d n^0_0 -{d\over dt}
\tilde n_0=0
\end{eqnarray}

{}From the homotopy formula (\ref{hf}) applied to $\tilde m$ 
and the cocycle condition, we get that $\tilde m=\tilde
m_0+s_H()-\tilde d(\int_0^1 {d\lambda\over\lambda}(\rho \tilde n_0)
[\lambda
z^\alpha])$ because $\rho$ (anti)commutes with $\tilde d$ and
$\tilde d$ is homogeneous of degree $0$ in $z^\alpha$. The last
expression vanishes since $\rho \tilde n_0=0$. Injecting the
remaining expression into the cocycle condition, we get 
\begin{eqnarray}
s_\omega \tilde m_0+\tilde d\tilde n_0=0. 
\end{eqnarray}

The homotopy formula (\ref{hf}) applied to $m^0$, together with
the cocycle condition, implies that 
\begin{eqnarray}
m^0 =m^0_0+ s_H()-\tilde d(\int_0^1 {d\lambda\over\lambda}(\rho
n^0_0)
+\int_0^1 {d\lambda\over\lambda}\rho \nonumber\\
(\sigma^{CA}\partial_{(k)}(-{\cal
L}_C-{\tilde\delta^R h\over\delta\tilde\phi^C}
-{\tilde\delta^R\over\delta\tilde\phi^C}
\{\tilde\phi^*_B\tilde\phi^B,\omega\}_{P,alt})
{\partial^L\over\partial(\partial_{(k)}\tilde\phi^A)}\tilde n_0)
[\lambda z^\alpha]\nonumber\\
=m^0_0+s_H()-\{\tilde\phi^*_A\tilde\phi^A,\tilde n_0
\}_{P,alt},
\end{eqnarray}
proving in particular the lemma on the dependence of $m$ on the
coordinates $z^\alpha$.
Injecting this last expression in the cocyle condition, 
using $s_\omega n^0_0=0$ and (\ref{com}), 
implies  
\begin{eqnarray}
s_\omega m^0_0+\tilde d n^0_0-{\partial\over\partial t}\tilde n_0+
\{h,\tilde n_0\}_{P,alt}=0.
\end{eqnarray}

\noindent {\bf Next steps in the lifting procedure}

We then have to try to lift the equivalent representative of $m$ given
by 
\begin{eqnarray}
m^\prime=dt (-\{\tilde\phi^*_A\tilde\phi^A,\tilde
n_0\}_{P,alt}+m^0_0)+\tilde m_0, \label{mp}
\end{eqnarray}
i.e., find $l=dt l^0+\tilde l$ such
that $s_H l+ dm^\prime=0$. This implies 
\begin{eqnarray}
s_H \tilde l+ \tilde d \tilde
m_0=0,\ s_H l^0 +\tilde d (-\{\tilde\phi^*_A\tilde\phi^A,\tilde
n_0\}_{P,alt}+m^0_0)-{d\over dt}\tilde m_0=0.
\end{eqnarray}
By exactly the same reasoning as before, the first equation implies
that 
\begin{eqnarray}
\tilde l=\tilde l_0+s_H(),\ 
s_\omega\tilde l_0+\tilde d
\tilde m_0=0.\label{sauv}
\end{eqnarray}
The second equation implies as before that 
\begin{eqnarray}
l^0=l^0_0+s_H()-\{\tilde\phi^*_A\tilde\phi^A,\tilde m_0
\}_{P,alt}
\end{eqnarray}
because $\rho$ annihilates the supplementary
$\tilde\phi^*_A$-dependent term which does not depend on ${\cal L}_A$. 
Injecting into the cocycle condition, we get 
\begin{eqnarray}
s_\omega l^0_0+
\tilde d m^0_0-{\partial\over\partial t}\tilde m_0+
\{h,\tilde m_0\}_{P,alt}=0,
\end{eqnarray}
the supplementary antifield dependent
term in $m^0$ cancelling the term coming from (\ref{com}) using the
fact that $s_\omega \tilde m_0+\tilde d\tilde n_0=0$. This shows that
at every step, we get the same dependence on the coordinates
$z^\alpha$, i.e, independence on $\partial_{(\mu)}{\cal
L}_A$, or by going back to the old coordinate system, on the time
derivatives of the fields, with a linear dependence in the antifields
and their spatial derivatives 
$\partial_{(k)}\tilde\phi^*_A$ only in the terms involving the
differential $dt$. Furthermore, we have proved the set of equations
(\ref{1})-(\ref{ch}). 

\noindent {\bf Coboundary condition}

Let us now consider the coboundary condition for $l^\prime$ defined in
an analoguous way as $m^\prime$ in (\ref{mp}). 
{}From $l^\prime=s_H r+d u$, we have, by applying $s_H$, that $s_H u + d
p=0$. Hence $u$ satisfies the same equation than $l$ above, which
implies by (\ref{sauv}) and an appropriate modification of $r$, that
we can assume $\tilde u=\tilde u_0$ and $u^0=u^0_0-
\{\tilde\phi^*_A\tilde\phi^A,\tilde p_0
\}_{P,alt}$. 

We have that
$\tilde l_0=s_H\tilde r+\tilde d \tilde u$, which implies, by applying the
homotopy formula (\ref{hf}) to $r$ 
that we can assume that $\tilde r=\tilde r_0$, $\tilde
u=\tilde u_0$. The coboundary condition becomes 
$\tilde l_0=s_\omega\tilde r_0+\tilde d \tilde u_0$, proving equation
(\ref{sch}).

By applying the homotopy formula (\ref{hf}) to $r^0$, 
the coboundary condition 
\begin{eqnarray}
-\{\tilde\phi^*_A\tilde\phi^A,\tilde
m_0\}_{P,alt}+l^0_0=-s_H r^0 -\tilde d u^0 +{d\over dt} \tilde
u_0, 
\end{eqnarray}
implies 
\begin{eqnarray}
-\{\tilde\phi^*_A\tilde\phi^A,\tilde
n_0\}_{P,alt}+l^0_0=-s_\omega r^0_0-
s_H\int_0^1 {d\lambda\over\lambda}(\rho(+\{\tilde\phi^*_A\tilde\phi^A,\tilde
m_0\}_{P,alt}\nonumber\\
-l^0_0-\tilde d u^0+{d\over dt} \tilde
u_0))[\lambda z^\alpha])-\tilde d u^0 +{d\over dt} \tilde u_0.
\end{eqnarray}
This gives
\begin{eqnarray}
l^0_0=-s_\omega r^0_0-\tilde d u^0_0 +{\partial\over \partial t} 
\tilde u_0-\{h,\tilde u_0\}_{P,alt},
\end{eqnarray}
proving equation (\ref{cob}).
These coboundary conditions are satisfied by choosing in the equation
$l=s_H r +du$, $r$ to be given by $dt(-\{\tilde\phi^*_A\tilde\phi^A,\tilde
u_0\}_{P,alt}+r^0_0)+\tilde r_0$ and a similar equation holding for $u$.
This proves (\ref{cob1}) \qed.
\vfill
\eject


\begin{thebibliography}{999}
\baselineskip=12truept
\bibitem{Zinn-Justin} J. Zinn-Justin, {\em Renormalisation of gauge 
theories} in {\em Trends in elementary particle theory}, Lecture 
Notes in Physics 
n$^0$37 (Springer, Berlin 1975)~; B.W. Lee, {\em Gauge theories} in
{\em Methods in field theory}, Les Houches Lectures 1975, eds. R.
Balian and J. Zinn-Justin (North-Holland, 1976)~; H. Kluberg-Stern 
and J.B. Zuber,
{\em Phys. Rev.} {\bf D12} (1975) 467, 482, 3159~; J.A. Dixon, Nucl.
Phys. {\bf B99} (1975) 420.
\bibitem{Batalin} I.A. Batalin and G.A. Vilkovisky, {\em 
Phys. Lett.} {\bf 102B} (1981) 27~;
{\em Phys. Rev.} {\bf D28} (1983) 2567~; {\em Phys. Rev.} 
{\bf D30} (1984) 508.
\bibitem{Bochicchio} M. Bochicchio, Phys. Lett. {\bf 193B} (1987) 31~;
C.B. Thorn, Phys. Rep. {\bf 175} (1989) 1~; B. Zwiebach, Nucl. Phys.
{\bf B390} (1993) 33~; 
E. Witten, Phys. Rev. {\bf D46} (1992) 5467~; E. Verlinde, Nucl. Phys.
{\bf B381} (1992) 141~; 
H. Hata and B. Zwiebach, Ann. Phys. {\bf 229} (1994) 177.
\bibitem{Dickey} L.A. Dickey, {\em Soliton equations and
Hamiltonian systems},
Advanced Series in Mathematical Physics, Vol. 12 
(World Scientific 1991).
\bibitem{BFV} E.S. Fradkin and G.A. Vilkovisky, {\em 
Phys. Lett.} {\bf 55B} (1975) 224~; I.A. Batalin and G.A. Vilkovisky,
{\em Phys. Lett.} {\bf 69B} (1977) 309~; E.S. Fradkin and
T.E. Fradkina, {\em Phys. Lett.} {72B} (1977) 343.
\bibitem{Henneaux} M. Henneaux and C. Teitelboim, {\em 
Quantization of Gauge Systems}, 
Princeton University Press (Princeton: 1992).
\bibitem{BBH} G. Barnich, F. Brandt and M. Henneaux,
{\em Local BRST cohomology in the antifield formalism: I. General 
theorems},
preprint ULB-TH-94/06, NIKHEF-H 94-13, hep-th 9405109, to appear in
Commun. Math. Phys.
\bibitem{Olver} P.J. Olver, 
{\em Applications of Lie Groups to Differential Equations}, 
Graduate Texts in Mathematics, volume 107, Springer Verlag 
(New York: 1986).
\bibitem{Witten} E. Witten, Mod. Phys. Lett. {\bf A5} (1990) 487.
\bibitem{DuboisViolette} M. Dubois-Violette, M. Henneaux, M. Talon 
and C.M. Viallet, {\em Phys. Lett.} {\bf 267B} (1991) 81.
\bibitem{Schomblond} G. Barnich, M. Henneaux and C. Schomblond,
{\em Phys. Rev.} {\bf D 44} (1991) R 939.
\bibitem{Fisch}J. Fisch and M. Henneaux, {\em Phys. Lett.} {\bf 226B}
(1989) 80~; W. Siegel, {\em
Int. J. Mod. Phys.} {\bf A 4} (1989) 3951.
\bibitem{Barnich} G. Barnich, {\em Mod. Phys. Lett.} {\bf A 9} (1994)
665.
\bibitem{McCloud} P. Mc Cloud, {Class. Quant. Grav.} {\bf 11} 
(1994) 567.
\bibitem{Saunders} D.J. Saunders, {\em The Geometry of Jet Bundles},
London Mathematical Society Lecture Note Series 142, 
Cambridge University Press (Cambridge 1989).
\bibitem{Anderson} I. M. Anderson, {\em The variational bicomplex}, Academic
Press (Boston : 1994)~; {\em Contemp. Math.} {\bf 132} (1992) 51.
\bibitem{DeDonder} Th. De Donder, {\em Th\'eorie invariantive du calcul des
variations},
Gauthier-Villars \'eds. (Paris: 1935).
 


\end{thebibliography}
\end{document}